\documentclass[%
 reprint,
superscriptaddress,
floatfix,
 amsmath,amssymb,
 aps,
 prx,
longbibliography
]{revtex4-2}

\usepackage{graphicx}
\usepackage{dcolumn}
\usepackage{bm}
\usepackage{placeins}
\usepackage{float}
\usepackage{hyperref}
\usepackage{placeins}
\usepackage{times}
\usepackage{physics}
\usepackage{gensymb}
\usepackage{xcolor}
\usepackage{float}
\usepackage{hyperref}
\usepackage{enumitem}
\hypersetup{
    colorlinks,
    citecolor=blue,
    filecolor=black,
    linkcolor=black,
    urlcolor=black
}
\usepackage[title,toc]{appendix}
\usepackage[T1]{fontenc}

\renewcommand{\thesubsection}{\arabic{subsection}.}

\begin{document} 
\title{A hybrid quantum network linking telecom-wavelength atomic and solid-state nodes}


\author
{\normalsize{Y. Chai,$^{1\dag}$ 
D. Ghoshal,$^{1,3\dag}$
N. P. Tiwari,$^{1,3}$ 
A. Kolar,$^{1}$ 
B. Pingault,$^{1, 4, 5}$
H. Bernien,$^{1, 2, 3\ast}$
and T. Zhong$^{1\ast}$}\\
\vspace{0.3cm}
\small{$^{1}$Pritzker School of Molecular Engineering, University of Chicago, Chicago, IL 60637, USA}\\
\small{$^{2}$Institute for Quantum Optics and Quantum Information, Austrian Academy of Sciences, 6020 Innsbruck, Austria}\\
\small{$^{3}$Institute for Experimental Physics, University of Innsbruck, 6020 Innsbruck, Austria}\\
\small{$^{4}$Materials Science Division, Argonne National Laboratory, Lemont, IL 60439, USA}\\
\small{$^{5}$Q-NEXT, Argonne National Laboratory, Lemont, IL 60439, USA}\\
\small{$^\dag$These authors contributed equally to this work.} \\
\small{$^\ast$To whom correspondence should be addressed; E-mail:~hannes.bernien@uibk.ac.at,~tzh@uchicago.edu}
}

\date{\today}


\begin{abstract}

Photonic links between disparate quantum technologies—such as photon sources, memories, processors, clocks, and sensors—are key to scaling quantum networks~\cite{2008_kimble_quantuminternet} and realizing a versatile quantum internet for secure quantum communication, distributed quantum computing, and entanglement-enhanced metrology~\cite{2018_Hanson_quantum_internet}. In practice, each technology is most suitably implemented on a different quantum platform; the substantial spectral mismatch between them~\cite{2011_Delft_QD_Rb, 2014_Wrachtrup_DBATT_Na, 2016_Rastelli_QD_Cs, 2019_Rolston_Ba_Rb, 2015_Köhl_QD_Yb, 2015_USTC_QD_NdVYO, 2025_Wolters_QD_Cs, 2017_ICFO_Rb_PrYSO}, along with scarce native telecom interfaces, thus poses a major bottleneck to achieving efficient interconnections over long distances~\cite{2018_Tamura_telecomfiber}. Here we demonstrate the first deployed two-node hybrid network that operates entirely in the telecom C-band. Our approach uses no quantum frequency conversion~\cite{1992_Huang_qfc, 2019_Strassman_qfc_noise, 2025_ICFO_qfc} or external filtering~\cite{2024_UCL_QD_Rb}; instead, we develop a neutral atom single photon source and a solid-state rare-earth quantum memory that both operate in previously unexplored telecom regimes with state-of-the-art performance. The source achieves a high single-photon purity at 46~kcps, and the memory a storage efficiency of 10.6$\%$ with high multimode capacity. We leverage the intrinsic tunability of both systems to optimize their spectral overlap and demonstrate microsecond-level storage and retrieval with a large time-bandwidth product. Moreover, we showcase real-world networking competencies such as support for multiplexing across 37 temporal modes and preservation of non-classicality over fibers of 10.6 km (metropolitan) and 49.2 km (laboratory). Our work establishes a backbone for telecom-native quantum repeater links and unlocks a path towards high-bandwidth, large-scale quantum networking.

\end{abstract}

\maketitle

\vspace{-5pt}
Hybrid quantum networks seek to interconnect distinct platforms to harness their complementary strengths~\cite{2009_Wallquist_HybridAOMsolidstate_review, 2015_Kurizki_Hybrid_review}. Though still nascent, previous studies have investigated slow-light effects~\cite{2011_Delft_QD_Rb, 2014_Wrachtrup_DBATT_Na, 2016_Rastelli_QD_Cs}, two-photon interference~\cite{2019_Rolston_Ba_Rb}, photon-mediated driving~\cite{2015_Köhl_QD_Yb}, photon storage and retrieval~\cite{2015_USTC_QD_NdVYO, 2025_Wolters_QD_Cs}, and photonic quantum-state transfer~\cite{2017_ICFO_Rb_PrYSO} between distinct quantum systems. However, these efforts remain constrained by incompatible optical transitions and non-telecom interfaces, limiting their deployability. While quantum frequency conversion~\cite{1992_Huang_qfc} offers tunability to bridge this gap, it introduces significant experimental complexity and suffers from imperfect efficiencies and added noise~\cite{2019_Strassman_qfc_noise, 2025_ICFO_qfc}. A more direct approach is to employ quantum systems with native telecom transitions and mutually compatible bandwidths, as exemplified by a recent proof-of-principle demonstration, albeit still relying on active pulse shaping and external filtering~\cite{2024_UCL_QD_Rb}.

\begin{figure*}[ht!]
\centering
\includegraphics{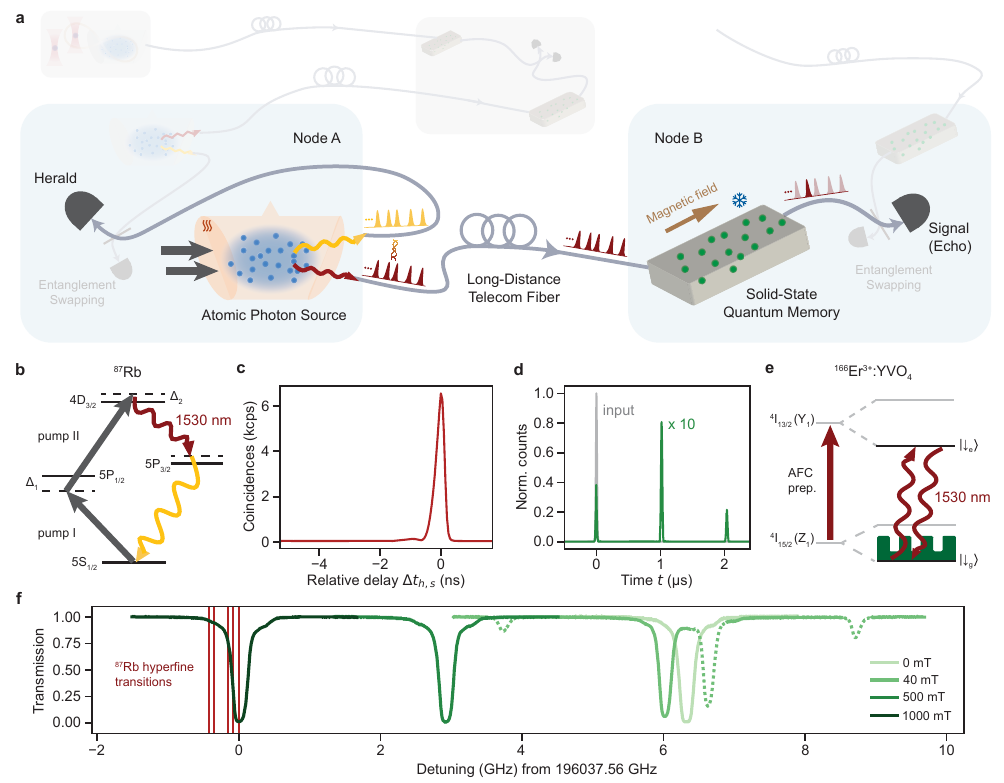}
\caption{\textbf{A hybrid two-node telecom quantum network.} \textbf{a}, An atomic single photon source at node A and a solid-state quantum memory at node B are connected through matched telecom photonic interfaces. This architecture forms a building block for a larger quantum network. \textbf{b}, Energy levels of $^{87}$Rb used for spontaneous four-wave mixing (4WM). The $|4D_{3/2}\rangle\rightarrow|5P_{3/2}\rangle$ telecom transition (maroon) is at 1530~nm, and the $|5P_{3/2}\rangle\rightarrow|5S_{1/2}\rangle$ transition (yellow) is at 780~nm. \textbf{c}, Coincidence histogram between the 780-nm heralding photon and 1530-nm signal photon, integrated over 20~s with a 50~ps time bin. \textbf{d}, Histogram of the storage and retrieval of a weak coherent pulse from an $^{166}$Er$^{3+}$:YVO$_{4}$ atomic frequency comb (AFC) quantum memory. The three peaks in green are the transmitted input pulse, the first-order, and the second-order photon echoes, respectively. \textbf{e}, Energy levels of $^{166}$Er$^{3+}$:YVO$_{4}$. Two crystal field levels are split into four Zeeman levels under an external magnetic field. The absorption line of the$|\downarrow_{g}\rangle\rightarrow|\downarrow_{e}\rangle$ transition at 1530~nm is used for AFC storage. \textbf{f}, $^{166}$Er$^{3+}$:YVO$_{4}$ absorption spectrum with respect to the hyperfine transitions of Rb. Spectrum at 40~mT shows all four Zeeman transitions, which are degenerate at zero magnetic field. The $|\downarrow_{g}\rangle\rightarrow|\downarrow_{e}\rangle$ transition (depicted by solid green lines) red-shifts with increasing magnetic fields. At around 1~T, it matches with the $^{87}$Rb hyperfine transitions~\cite{2009_Moon_Rb4Dhyperfine}. Dotted lines represent other optical transitions between the $|\downarrow_{g}\rangle, |\uparrow_{g}\rangle, |\downarrow_{e}\rangle, |\uparrow_{e}\rangle$ levels.}
\label{fig:architecture}
\end{figure*}

Among candidate platforms for hybrid network nodes, warm atomic vapors, realizing the spontaneous four-wave mixing (4WM) process~\cite{2006_Chaneliere_4WM}, provide bright, high-fidelity entangled photon sources that are exceptionally compact and robust~\cite{2021_Davidson_4WM, 2022_Kim_4WM, 2024_Craddock_4WM, 2020_Park_entswap, 2024_Craddock_4WMdeployed}. They are also naturally compatible with other atom-based systems such as quantum processors~\cite{2010_Saffman_rydberg} and clocks~\cite{2015_Ludlow_clock}, providing a straightforward route to integrating multiple functionalities into the network. Rare-earth-ion doped crystals, meanwhile, are a leading technology in optical quantum memories, offering excellent optical and spin coherence, together with broadband operation arising from ensemble inhomogeneous broadening. The atomic frequency comb (AFC) memory protocol~\cite{2009_Gisin_AFC_OGTh, 2008_Gisin_AFC_OGExp} supports storage of photons in multiple temporal, spectral~\cite{2014_Simon_Oblak_AFCmultimode}, and spatial modes~\cite{2025_ICFO_AFCspmultimode, 2025_USTC_AFCspmultimode}. With a judicious choice of atomic species, native telecom-band operation of both a high-performance photon source and quantum memory would enable their direct connection as an efficient elementary link of a quantum repeater (Fig.~\ref{fig:architecture}a), in which long-distance entanglement distribution is divided into shorter segments interconnected via entanglement swapping~\cite{1998_Briegel_repeater, 2023_Azuma_repeater}. Such repeater links would greatly boost entanglement rates given the high capacity of the quantum memory~\cite{2025_IQOQI_ICFO_Hybrid, 2025_Sorensen_Hybrid_1, 2025_Borregaard_Tm_Rbcavity_th}, and allow incorporation of quantum processors, notably atoms trapped in optical tweezer arrays~\cite{2024_Lukin_atomarray}, thus opening new frameworks for hybrid quantum information processing~\cite{2021_Sangouard_qubit_multimodememory, 2017_Schuster_Multimode}.

Here, we construct a two-node network comprised of a tunable atomic photon source and solid-state quantum memory, whose precise spectral alignment in the telecom band readily enables critical capabilities: multiplexed interconnection and metropolitan-scale deployment. The two nodes are located in separate laboratories and directly connected by optical fibers at 1530~nm. Leveraging the intrinsic tunability of each system, we achieve spectral matching with 100-MHz bandwidth at the single-photon level without frequency conversion or external filtering. We employ a Rb atomic vapor as a high-purity heralded single-photon source and demonstrate microsecond-level storage and retrieval in an Er-based quantum memory. By jointly optimizing the two systems, we achieve a rate of up to 4.3(1)~cps while preserving nonclassical correlations. We further demonstrate the multimodality of the source–memory network and extend the link to 49.2~km in a laboratory setting. Finally, we integrate these capabilities to realize temporally multiplexed networking between the atomic and solid-state nodes via a 10.6-km fiber loop deployed across the Chicago metropolitan area, establishing a scalable foundation for hybrid quantum networks.

\begin{figure*}[ht!]
\centering
\includegraphics{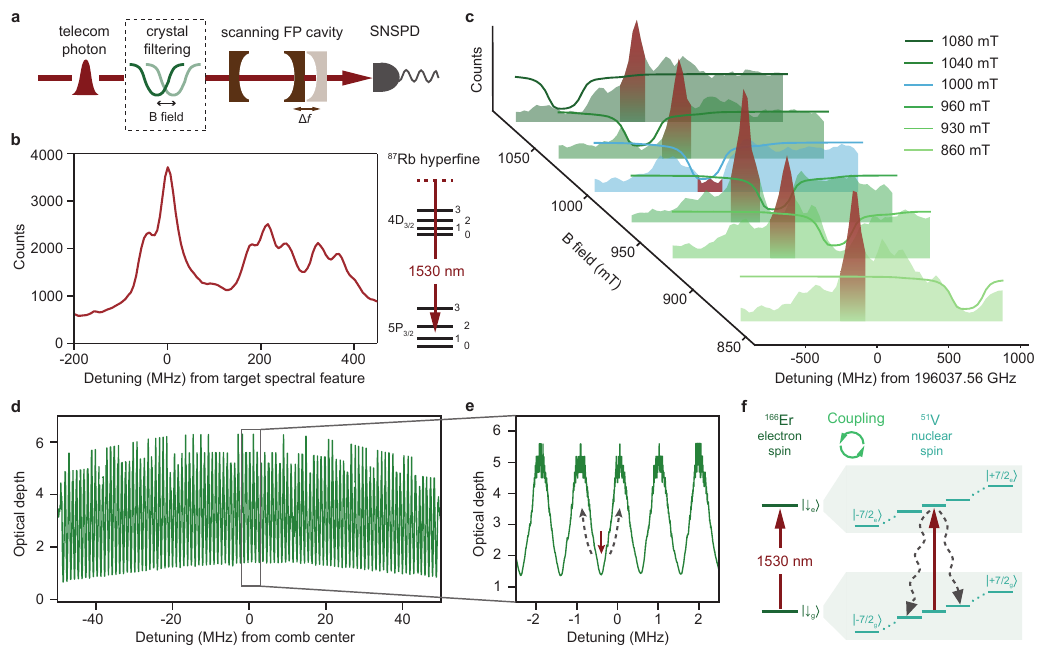}
\caption{\textbf{Matching telecom photonic interfaces.} \textbf{a}, Single-photon spectrometer. The 1530-nm photon from the source is directed through a scanning Fabry-P\'erot cavity and to an SNSPD. Spectra without and with crystal absorption as a notch filter are shown in \textbf{b} and \textbf{c}, respectively. \textbf{b}, Source node 1530-nm photon spectrum with blue-detuned two-photon pumping. The hyperfine levels involved are shown on the right. \textbf{c}, 1530-nm photon spectra with crystal filtering. At various magnetic fields, the crystal transmission (solid lines) and corresponding transmitted photon spectra (translucent green shades) are overlaid. At 1000~mT (blue), the target spectral feature (red) is absorbed by the crystal, indicating the spectral match of the two systems. All spectra are frequency-referenced to a calibrated wavemeter. \textbf{d}, Absorption spectrum of a 100-MHz wide atomic frequency comb (AFC) at the matching frequency identified in \textbf{c}. \textbf{e}, Zoom-in spectrum near the center of the AFC, showing a 1-MHz comb spacing. \textbf{f}, Super-hyperfine levels involved in the spectral hole-burning. Arrows in \textbf{e} indicate the population transfer responsible for spectral holes.}
\label{fig:interface}
\end{figure*}


\subsection*{\texorpdfstring{$^{87}$Rb vapor and $^{166}$Er$^{3+}$:YVO$_{4}$ crystal}{87Rb vapor and 166Er:YVO4 crystal}}
\vspace{-5pt}
Our atomic single photon source in node A is realized by spontaneous 4WM with a diamond level structure~\cite{2008_Rolston_diamond4WM, 2010_Rolston_diamond4WM, 2011_Rolston_polarization} in a 10-mm-long enriched $^{87}$Rb vapor cell heated to 92$^{\circ}$C. We use continuous-wave 795-nm and 1475-nm lasers to address the two-photon transition $|5S_{1/2}\rangle \rightarrow |4D_{3/2}\rangle$ via the intermediate $|5P_{1/2}\rangle$ state (Fig.~\ref{fig:architecture}b). Cascaded decay back to the ground state through another intermediate state $|5P_{3/2}\rangle$ generates a pair of correlated photons, one at 1530~nm (signal photon) and the other at 780~nm (heralding photon). We use a collinear pump geometry with focusing lenses, inspired by~\cite{2024_Craddock_4WM}, to fulfill the phase matching condition for efficient photon pair generation. The coincidence histogram in Fig.~\ref{fig:architecture}c shows a bi-photon detection rate of 46~kcps with a second-order cross-correlation $[g^{(2)}_{h,s}]_{\mathrm{max}} =$~130(5), which represents the state-of-the-art for an atomic photon pair source in the telecom C-band. The heralded 1530-nm photon has a temporal correlation time of 0.32(2)~ns. The temporal and spectral profile of the heralded 1530-nm photons are widely tunable using different pump powers, intermediate detunings $\Delta_{1}$, and two-photon detunings $\Delta_{2}$. We harness this tunability to optimize the spectral overlap with the memory (Supplementary Information). 

Our solid-state quantum memory in node B is based on an optical AFC~\cite{2008_Gisin_AFC_OGExp, 2009_Gisin_AFC_OGTh} in an isotopically purified $^{166}$Er$^{3+}$:YVO$_{4}$ crystal with 15 parts per million (ppm) erbium doping concentration. We choose YVO$_{4}$ as the host matrix for erbium dopants because their optical transition between the $|^{4}I_{15/2},Z_{1}\rangle$ and $|^{4}I_{13/2}, Y_{1}\rangle$ crystal field levels (Fig.~\ref{fig:architecture}e) is at 1530~nm (Methods), and can be tuned to the exact Rb resonance with a magnetic field. We prepare an AFC by performing spectral hole-burning on the optical transition between the two lower Zeeman levels (i.e. $|\downarrow_{g}\rangle$ and $|\downarrow_{e}\rangle$) of the $^{166}$Er$^{3+}$ ensembles. A resonant photon absorbed by an AFC with a comb spacing $\Delta_{\mathrm{AFC}}$ leads to an echo emission after a storage time of $\tau_{\mathrm{AFC}}=\frac{1}{\Delta_{\mathrm{AFC}}}$.
A histogram for the optical AFC storage in $^{166}$Er$^{3+}$:YVO$_{4}$ is shown in Fig.~\ref{fig:architecture}d, where we measure the first-order AFC echo at a delay of 1.0~$\mu$s with 10.6(1)$\%$ storage efficiency for a weak coherent input pulse. The storage time can be prolonged up to 3~$\mu$s without significantly degrading the efficiency (Supplementary Information).

With both the source and the memory node operating at 1530~nm, we examine the spectral proximity between the two via optical spectroscopy of the $|^{4}I_{15/2}, Z_{1}\rangle \rightarrow |^{4}I_{13/2}, Y_{1}\rangle$ transition in the crystal (Fig.~\ref{fig:architecture}f). At zero magnetic field, the memory transition is blue-detuned by 6.4-6.8~GHz from the $^{87}$Rb hyperfine transitions. Owing to the difference between the excited- and ground-state electron-spin $g$-factors of Er in YVO$_{4}$ ($g_e$~=~4.51, $g_g$~=~3.54~\cite{2021_Faraon_ErYVO}), increasing magnetic field along the crystal $c$-axis red-shifts the $|\downarrow_{g}\rangle \rightarrow |\downarrow_{e}\rangle$ transition. At $B\approx$~1~T, this Er transition overlaps with the Rb transitions, and has an inhomogeneously broadened linewidth of $\Gamma$ = 131(1)~MHz, which sets the maximum memory bandwidth to absorb signal photons from the source. These results confirm that the atomic photon source and solid-state quantum memory are spectrally aligned and tunable for precise telecom-interface matching.

\subsection*{Matching telecom photonic interfaces}
\vspace{-5pt}
Next, we refine the spectral matching to MHz level. Although the photon spectral bandwidth can be inferred from the coincidence histogram in Fig.~\ref{fig:architecture}c, efficient storage in a quantum memory requires detailed knowledge of the full spectral profile. We therefore explicitly measure the 1530-nm photon spectrum via high-resolution spectroscopy (Fig.~\ref{fig:interface}a) with a scanning Fabry-P\'erot (FP) cavity and a superconducting nanowire single-photon detector (SNSPD). Figure~\ref{fig:interface}b shows the spectrum with pump detunings $\Delta_{1}=-$817~MHz and $\Delta_{2}=+$903~MHz, which reveals significantly richer structure than the coincidence measurements alone. This is, to our knowledge, the first measurement for such an atomic source. Three groups of spectral features are observed over a range $\gtrsim$~600~MHz, arising from a combination of decay paths through different hyperfine levels and the distribution of participating atomic velocity classes. The absolute frequencies and relative intensities of these features can be fine-tuned by adjusting $\Delta_{1}$ and $\Delta_{2}$ (Supplementary Information). Here, we select a regime that optimizes the overall source performance while minimizing the magnetic field tuning for the crystal. The spectral feature centered at 0~MHz has the highest intensity with a bandwidth of about 100~MHz, and is chosen as the target for matching. To find the optimal magnetic field for the crystal, we introduce the crystal as a tunable notch filter for the 1530-nm photons from Rb prior to the scanning FP cavity (Fig.~\ref{fig:interface}a): it absorbs a spectral band of width $\Gamma$ while transmitting the remaining components. The results of this fine spectral matching procedure are plotted in Fig.~\ref{fig:interface}c. The absorbed band in the photon spectrum (green shade) red-shifts with increasing magnetic field, coinciding with the independently measured Er absorption profile (green line). At $B=$~1000~mT, the target spectral feature completely disappears, confirming its efficient absorption by the crystal.

While maintaining the spectral matching at a fixed magnetic field, we prepare an optical AFC in $^{166}$Er$^{3+}$:YVO$_{4}$ that simultaneously achieves high bandwidth (100~MHz), high efficiency and high capacity. As shown in Fig.~\ref{fig:interface}d, we create an AFC comb by repeating spectral hole-burning at 100 discrete frequencies with a spacing $\Delta_{\mathrm{AFC}}=1$~MHz across a total absorption window of $\Gamma_{\mathrm{AFC}}=100$~MHz. We realize a state-of-the-art optical AFC memory in the telecom-C band with an optimized storage efficiency of 7.7(1)$\%$ for a weak coherent pulse with a full-width-at-half-maximum (FWHM) bandwidth of 43~MHz and higher efficiency for narrower inputs (Supplementary Information). Previously, such performance has been proven difficult due to relatively inefficient hole-burning with Er electron spins~\cite{2011_Gisin_ErAFC, 2016_Oblak_Tittel_ErAFC, 2020_Oblak_Tittel_ErAFC}. AFCs based on long-lived hyperfine levels of $^{167}$Er$^{3+}$ showed significant improvements~\cite{2021_Faraon_ErAFC, 2021_Sellars_ErAFC, 2022_Guo_ErAFC}, but these demonstrations so far are exclusive to $^{167}$Er$^{3+}$:Y$_{2}$SiO$_{5}$ crystals operating at 1539~nm. Our AFC uses neither the electron nor the nuclear spins of Er. We instead use the nuclear spins of vanadium (I$_V$=$\frac{7}{2}$) in the YVO$_{4}$ host. While the electron spin of $^{166}$Er$^{3+}$ is completely frozen at $B=$~1000~mT and an effective crystal temperature of 150~mK (estimated from Zeeman level populations), the super-hyperfine couplings between an $^{166}$Er$^{3+}$ and its neighboring $^{51}$V$^{5+}$ spins offers a band of closely-spaced sub-levels within the Er ground $|\downarrow_{g}\rangle$ and excited state $|\downarrow_{e}\rangle$ (Supplementary Information). As a result, hole-burning redistributes population among neighboring ground-state nuclear-spin levels (Fig.~\ref{fig:interface}f), producing long-lived spectral holes with associated anti-holes distributed within 1~MHz of the central hole feature (Fig.~\ref{fig:interface}e). While the detailed hole-burning dynamics is a subject of future investigations, this mechanism facilitates the creation of a broadband AFC with dense teeth and high optical depths.

\begin{figure}[ht!]
\centering
\includegraphics{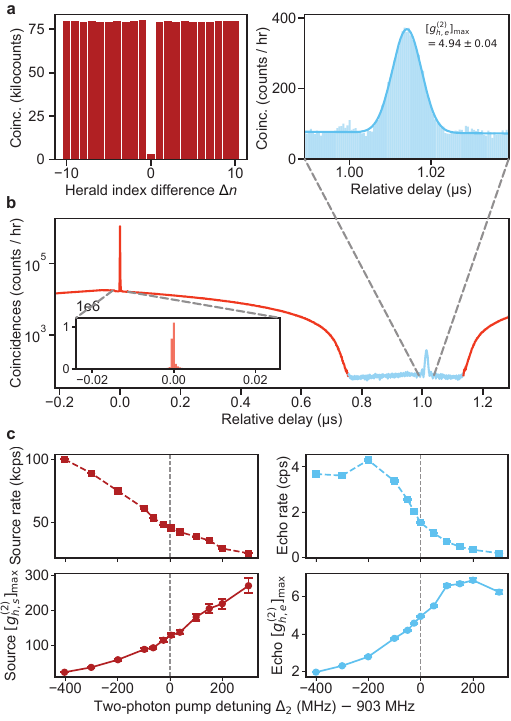}
\caption{\textbf{Source--Memory networking: single-photon storage and retrieval.} \textbf{a}, Hanbury-Brown-Twiss (HBT) interferometry: coincidence histogram between 1530-nm photons, heralded by a 780-nm photon detection, integrated for 600~seconds. For $\Delta n =$~0, the normalized heralded auto-correlation $g^{(2)}_{s,s|h}(0)=$~0.031(1). \textbf{b}, Coincidence histogram between the heralding and signal photons after storage and retrieval from the memory. The inset plots the direct transmission peak (orange) at 0~$\mu$s, and the outset shows the echo (blue) at 1.01~$\mu$s. The background noise at echo retrieval is minimized with a temporal gating scheme. Counts are integrated for 3 hours with a 0.5~ns time bin. \textbf{c}, Trade-off between heralded rate (top) and cross-correlation (bottom) of source photon (red) and echo photon (blue) with changing two-photon pump detunings $\Delta_{2}$ from $+$500~MHz to $+$1.2~GHz. The values and uncertainties are extracted from fitting. The vertical line corresponds to $\Delta_{2} =+$903~MHz used in \textbf{a} and \textbf{b} where the heralded echo rate is $R_{h,e}=$~1.5(1)~cps and the cross-correlation $[g^{(2)}_{h,e}]_{\mathrm{max}} =$~4.94(4). The highest echo rate $R_{h,e}=$~4.3(1)~cps occurs at $\Delta_{2} = +$703~MHz with a $[g^{(2)}_{h,e}]_{\mathrm{max}} =$~2.78(2).}
\label{fig:network}
\end{figure}

\begin{figure*}[ht!]
\centering
\includegraphics{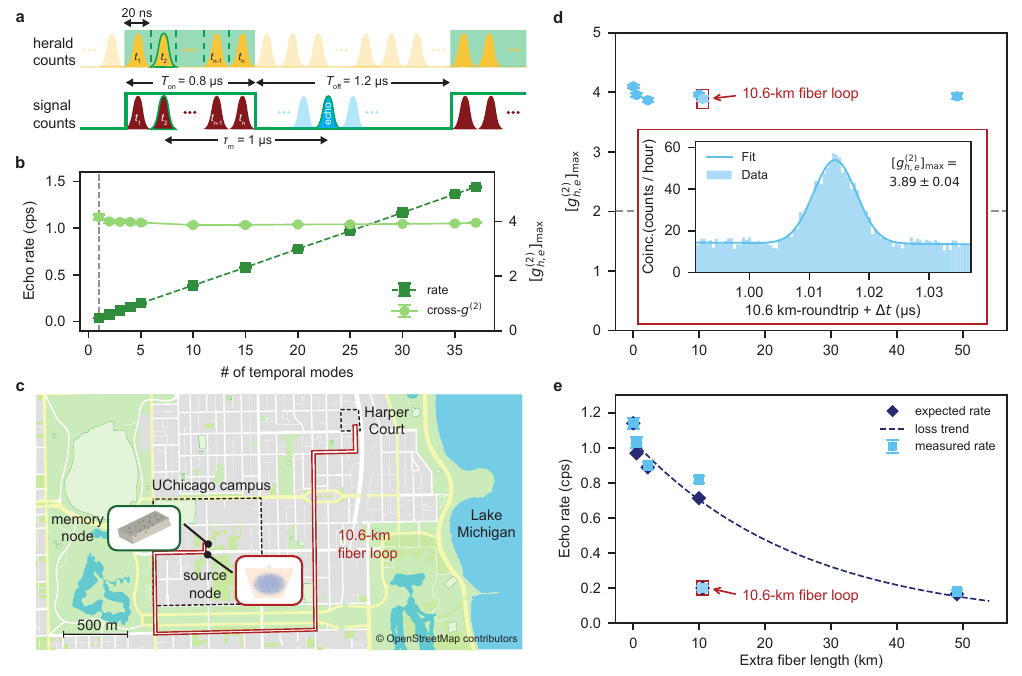}
\caption{\textbf{Multimode networking and metropolitan deployment.} \textbf{a}, Multiplexing scheme with a stream of continuously generated photon pairs. The maximal number of temporal modes for storage and retrieval increases with memory acceptance window (green). \textbf{b}, Heralded echo rate and cross-correlation $[g^{(2)}_{h,e}]_{\mathrm{max}}$ with varying mode numbers. 37 temporal modes are demonstrated with a memory acceptance window of 0.74~$\mu$s, without any degradation of the the cross-correlation. \textbf{c}, Map of the deployed 10.6-km fiber loop routed to and back from an off-campus location (Harper Court) in Hyde Park, Chicago. \textbf{d}, Cross-correlation $[g^{(2)}_{h,e}]_{\mathrm{max}}$ and \textbf{e}, Heralded echo rate with extended fibers up to 49.2~km (laboratory) and 10.6~km (metropolitan) inserted between the source and the memory node. Expected rates (navy) in \textbf{d} are inferred from the measured rate with no extra fiber and the independently measured fiber losses. Dashed line shows the expected loss with distance. The inset of \textbf{d} shows the zoomed-in echo coincidence for the 10.6-km deployed fiber loop; the corresponding data points are highlighted in red boxes in \textbf{d} and \textbf{e}.}
\label{fig:apply}
\end{figure*}

Together, these results constitute the first observation of spectral matching between two distinct functional nodes in the telecom C-band without employing quantum frequency conversion or pulse shaping.

\subsection*{Single-photon storage and retrieval}
\vspace{-5pt}
We demonstrate the hybrid network link by directly storing the 1530-nm single photons from the source node in the memory node. As a prerequisite for a quantum network, we first confirm operation in the single-photon regime by performing heralded Hanbury-Brown-Twiss (HBT) interferometry on the source 1530-nm photons~\cite{2004_Geneva_autog2}. The heralded auto-correlation $g^{(2)}_{s,s|h}(0)$ of the 1530-nm photon is minimized with large pump detunings (Supplementary Information). As shown in Fig.~\ref{fig:network}a, in the pump regime we operate in Fig.~\ref{fig:interface} and Fig.~\ref{fig:network}b ($\Delta_{1}=-$817~MHz, $\Delta_{2}=+$903~MHz) the photon exhibits strong anti-bunching with $g^{(2)}_{s,s|h}(0) =$~0.031(1), well below the two-photon Fock state bound of $g^{(2)}(0)\leq$~0.5. Notably, all measurements in this work are with modest pump powers ($\lesssim$~15~mW) and a low mean photon pair number of 0.007, leaving ample room in future work for increasing the source rates while maintaining high single-photon purity.

We then measure the coincidences between the heralding photons and the stored signal photons after retrieval from the memory. As in Fig.~\ref{fig:network}b, we measure at zero relative delay a coincidence peak that corresponds to the signal photons that are directly transmitted through the crystal without being stored. After a storage time of $\tau_{m}=\tau_{\mathrm{AFC}}=$~1.01~$\mu$s, we measure a second coincidence peak from stored-and-retrieved photons. To minimize background noise, we implement a temporal gating scheme in which the memory acceptance window is divided into alternating on–off blocks so that the source photon transmission is switched off at the expected retrieval~\cite{2017_ICFO_Rb_PrYSO} (Methods). This creates a low background at the retrieved photon echo in Fig.~\ref{fig:network}b, recovering a high signal-to-noise ratio (SNR) and a normalized non-classical correlation $[g^{(2)}_{h,e}]_{\mathrm{max}}=$~4.94(4), which is well above the threshold of 2 given by the Cauchy-Schwarz inequality~\cite{1986_Walls_CauchySchwarz}.

To benchmark our system's performance, we use the metric time-bandwidth product: TBP $=\tau_{m}/\Delta \tau_{e}$, where $\Delta \tau_{e}$ is the temporal FWHM of the echo coincidence~\cite{ 2016_Oblak_Tittel_ErAFC}. The temporal broadening of the echo ($\Delta \tau_{e}=$~8.4(1)~ns) compared to that of the pre-storage signal (0.32(2)~ns, Fig.~\ref{fig:architecture}c) is consistent with the frequency-selective storage of the photons fixed by the memory bandwidth $\Gamma_{\mathrm{AFC}}$. We obtain a TBP~=~120(1), which is one to two orders of magnitude higher than previous hybrid networking experiments~\cite{2017_ICFO_Rb_PrYSO, 2025_Wolters_QD_Cs, 2024_UCL_QD_Rb} and can be further enhanced by reducing $\Delta_{\mathrm{AFC}}$ while maintaining $\Gamma_{\mathrm{AFC}}$. The large TBP establishes a hybrid two-node network for highly multiplexed operation, which is investigated in the next section.

Leveraging the spectral tunability of the 1530-nm photons discussed previously, we further optimize the system performance by fine-tuning the two-photon pump detunings $\Delta_{2}$. The right panel of Fig.~\ref{fig:network}c shows a trade-off between the heralded echo rate and the maximum cross-correlation $[g^{(2)}_{h,e}]_{\mathrm{max}}$. The trade-off generally follows that of the source (Fig.~\ref{fig:network}c, left panel), with noticeable deviations at larger detunings due to the imposed spectral matching conditions. By analyzing the coincidence peaks before and after storage and retrieval, and taking into account the change in the photon temporal profile, we estimate an excess noise of 0.40(7)$\times$10$^{-3}$~cps, which includes and is dominated by the SNSPD dark counts, thus verifying negligible noise added by the memory (Supplementary Information). Across the entire fine-tuning range, we maintain a single-photon $g^{(2)}_{s,s|h}(0) <$~0.5 (Supplementary Information). At the highest internal storage efficiency of 0.53$\%$ (Methods), our hybrid source--memory link operates at an overall rate of 4.3(1)~cps while maintaining a non-classical $[g^{(2)}_{h,e}]_{\mathrm{max}}$.

\subsection*{Multiplexing and field-test with deployed fibers}
\vspace{-5pt}
Building on the large TBP and the fine-tuned operating points established above, our hybrid source--memory link enables high temporal multiplexing capabilities with straightforward implementation. Our photon source, operated under continuous-wave pump, generates photon streams without temporal or physical overhead from control pulses or system resets. On the other hand, our quantum memory allows for storage and retrieval of continuous streams of photons for a duration up to $T_{on}$, the on-block within the memory acceptance window with a constraint $T_{on}<\tau_{m}$ in our gating scheme (Supplementary Information). Within this interval, up to $N=T_{on} / \Delta t_{e}$ temporal modes can be stored simultaneously (Fig.~\ref{fig:apply}a), where $\Delta t_e$ ($\gtrsim 2\Delta \tau_{e}$) is the full duration of a single photon echo. From our echo coincidence histogram (Fig.~\ref{fig:network}b), we extract $\Delta t_e=$~20~ns, which captures 99.5$\%$ of the heralded echo counts. Increasing $T_{on}$ in increments of $\Delta t_e$ correspondingly increases the number of supported temporal modes. Figure~\ref{fig:apply}b displays a linear increase of the heralded echo rate with the number of temporal modes, accompanied by the preservation of $[g^{(2)}_{h,e}]_{\mathrm{max}}$ across all modes. Our hybrid network link operates with up to 37 independent temporal modes while maintaining a cross-$g^{(2)}$ above the classical limit of 2. Such temporal multiplexing is highly advantageous in long-distance entanglement distribution, as it allows for multiple entanglement generation attempts within the classical communication time between adjacent links, and prepares for future network applications such as synchronization and buffering~\cite{2014_Simon_Oblak_AFCmultimode, 2025_IQOQI_ICFO_Hybrid, 2025_Sorensen_Hybrid_1, 2025_Borregaard_Tm_Rbcavity_th}.

Finally, we showcase multiplexed single photon storage and retrieval of our hybrid network with extended fiber distances between the source and memory nodes. We measure the cross-correlation ($[g^{(2)}_{h,e}]_{\mathrm{max}}$) and the heralded echo rate for various lengths of spooled fibers, as well as for a 10.6-km deployed fiber loop in the Chicago metropolitan area (Fig.~\ref{fig:apply}c). No further experimental complexity is necessary other than electronically compensating for the relative delays of the 1530-nm photons from the additional traveling distances. As shown in Fig.~\ref{fig:apply}d and \ref{fig:apply}e, with extended fiber spools the measured echo rate aligns with the nominal fiber attenuation of 0.20-0.35~dB/km at 1530~nm, while $[g^{(2)}_{h,e}]_{\mathrm{max}}$ remains at the same non-classical level for all distances, showing no degradation. For the 10.6-km field test, we assemble a fiber loop stretching across the Hyde Park neighborhood from the laboratories at the University of Chicago to the one at Harper Court. The loop includes several additional fiber splices and fiber-to-fiber connections, resulting in a higher attenuation of 7.56~dB compared to a fiber spool with the same length. Nevertheless, the deployed fiber loop markedly maintains the identical $[g^{(2)}_{h,e}]_{\mathrm{max}}$, therefore, does not add excess noise to the network even without filtering or active stabilization. The 1530-nm single photons are successfully retrieved at a rate of 0.20(1)~cps, with a cross-correlation $[g^{(2)}_{h,e}]_{\mathrm{max}}=$~3.89(4), demonstrating the performance of the hybrid network under realistic fiber-network environment.

\section*{Discussion}

The development of a large-scale quantum network relies on the realization of functional quantum nodes and their photonic interfaces to existing telecom infrastructures. The hybrid source--memory network presented in this work links two quantum systems with significantly different characteristics directly in the telecom C-band. Each system has its own tunability that can be independently utilized to optimize the spectral matching condition as well as to enable multiplexing, offering a boost for the success rate. The established link represents a building block for a large-scale hybrid quantum network and has shown robustness in a real-world metropolitan setting with long-distance deployed fibers.

Toward a full hybrid quantum network, our work lays the foundation for realizing remote entanglement and quantum-state transfer between the atomic photon source and the solid-state quantum memory. The atomic photon pair source can be configured to generate hyper-entanglement with both time-energy and polarization encoding. The former is inherent in a continuous-wave pumped pair source with multiple entangled time bins within the coherence time of the pump field~\cite{2015_Zhong_timeenergy},  whereas the latter can be attained by isolating Zeeman sub-levels in the hyperfine state manifolds~\cite{2024_Craddock_4WM, 2024_Craddock_4WMdeployed, 2011_Rolston_polarization}. The solid-state AFC memory is naturally compatible with time-energy entanglement~\cite{2021_ICFO_AFCinterferometer}, and can be adapted for polarization-entangled photons by converting them to spatial-mode entanglement and storing them in separate regions of the crystal \cite{2025_ICFO_AFCspmultimode, 2025_USTC_AFCspmultimode}. Such multiplexed, hyper-entanglement offers versatile resource states in a hybrid quantum network, potentially enabling high-speed teleportation between the source and memory nodes. While the optical AFC memory in this work has pre-determined storage time, the long-lived, multi-level vanadium nuclear spins in YVO$_4$ provide a plausible realization of spin-wave AFC~\cite{2021_ICFO_AFCinterferometer, 2010_Gisin_AFCspinwave} in the telecom band with on-demand retrieval. A full quantum repeater can be realized by duplicating the source--memory entanglement link into two pairs of nodes.

Moreover, the native 1530-nm atomic transition of Rb allows for direct telecom photonic interfaces for Rb atomic qubits trapped in arrays of optical tweezers and coupled to optical cavities~\cite{2023_Covey_atomcavityreview, 2021_Lukin_Rbcavity, 2025_Borregaard_Tm_Rbcavity_th, 2025_Lukin_Rbcavity}. A high-speed telecom entanglement link between a Rb quantum processor and a high-capacity solid-state Er quantum memory will unlock new paradigms of hybrid quantum information processing with potential advantages of reduced processor qubit counts and faster computation~\cite{2021_Sangouard_qubit_multimodememory}. The Er memories can also serve to interconnect, synchronize and multiplex distributed modular atomic processors over telecom fibers. Together, this new architecture opens up a promising approach to scaling quantum computing networks.



\renewcommand{\appendixname}{Methods}
\section*{Methods}
\subsection*{Experimental setup}
\vspace{-5pt}
Figure~\ref{fig:setup} provides a detailed schematic of the experimental setup. A laser module prepares all the required optical fields: two 4WM pump lasers for Node A (photon source node) and one AFC preparation laser for Node B (quantum memory node). The stabilization and monitoring of each laser is shown in Fig.~\ref{fig:setup}a and discussed in the following sub-sections respectively.

There are three detector modules carrying out different types of measurements discussed in the main text. All single-photon level detection is done using a multi-channel SNSPD (Quantum Opus) together with a Time Tagger (Swabian Instruments). For the telecom photon spectra measurements, a small fraction of the same comb preparation laser is sent through the scanning FP cavity so that we can use the corresponding signal with the frequency reading from the wavemeter as a frequency reference for all spectra data.

The two quantum nodes are located in two laboratories separated by 35 meters. All detector modules are in the same laboratory with node A. Two 50-m steel-jacketed optical fibers are deployed for routing photons between the two nodes and to the detector modules.

\subsection*{Experimental control}
\vspace{-5pt}
Figure~\ref{fig:control} provides an overview of the experimental control sequence. All control pulses are generated from an arbitrary waveform generator (HDAWG, Zurich Instruments) located in the same laboratory as node B. The switching between AFC preparation and storage is implemented through an optical MEMS switch (Sercalo SW1$\times$2-9N). For one full sequence, AFC preparation takes 3.84~s. After the 0.5~s wait time, the storage window opens for 4~s. There is a 43~ms rest time at the end of the sequence before the next cycle begins. At the beginning of the storage acceptance window, a trigger is sent from the HDAWG to the Time Tagger to start counting. The entire pulse sequence takes 8.4~s, with an active storage duty cycle of $47.6\%$. A second optical MEMS switch (Sercalo SW1$\times$2-9N) is used as a shutter in the detection path to block the AFC preparation laser from transmission to the SNSPD, and is only on during the AFC storage acceptance window.

\subsection*{Atomic photon source}
\vspace{-5pt}
The atomic photon pair source at node A is based on a custom cylindrical rubidium vapor cell (Precision Glassblowing) with $>98\%$ isotopically purified $^{87}$Rb. The cell is 10-mm long, with wedged windows attached at an angle to minimize back-reflections. The cell is heated to $92\degree$C with ceramic ring resistive heaters on both windows, with a constant applied current. Custom thermal insulation--- aluminum foil to reflect thermal radiation back into the cell, a fiberglass blanket, and a fiberglass post for mounting---keep the temperature stable to $<0.5\degree$C without active stabilization.

We use 4WM with a diamond level structure to generate the correlated photon pairs (Fig.~\ref{fig:architecture}b). We apply two continuous-wave pump lasers at 795~nm and 1475~nm to address the two-photon transition $|5S_{1/2},F=2\rangle \rightarrow |5P_{1/2},F'=1\rangle \rightarrow|4D_{3/2}\rangle$, addressing all allowed hyperfine levels in $|4D_{3/2}\rangle$ due to Doppler shifts. We collect correlated photons at 1530~nm and 780~nm generated through the path $|4D_{3/2}\rangle \rightarrow |5P_{3/2}\rangle \rightarrow|5S_{1/2}\rangle$.

The first laser at 795~nm is a pigtailed distributed Bragg reflector single-frequency laser (Thorlabs DBR795PN) and is frequency-monitored using saturated absorption spectroscopy. The second laser at 1475~nm is an external cavity diode laser (Toptica DL pro) and is actively frequency-stabilized to within 1 MHz via a multi-channel wavemeter (HighFinesse WS7) and PID logic on a computer. Each pump is power-stabilized to within 1\% using a half-wave plate on a motorized rotation mount, PBS, beam sampler, photodiode, and logic on a computer (Fig.~\ref{fig:setup}b).

For efficient 4WM photon pair generation, the four beams must satisfy the phase-matching condition $\vec{k}_{795} + \vec{k}_{1475} = \vec{k}_{1530} + \vec{k}_{780}$. Our optical layout using a collinear geometry (Fig.~\ref{fig:setup}b) is inspired by~\cite{2024_Craddock_4WM}. The two pump beams are combined before the cell using a dichroic mirror. After the cell, the generated beams are separated from each other and isolated from the pumps using another dichroic and multiple stacked narrow-line optical filters. In between the two dichroics we use a pair of achromatic $f =100$~mm lenses to focus both pump beams to a $1/e^{2}$ diameter of $\approx 95~\mu$m inside the cell. 

Fiber coupling correlated 780-nm and 1530-nm photons requires some care, as the beams have small spatial modes (on the order of that of the pump beams) and are invisible to IR cards or photodiodes typically used for free-space alignment. Due to our collinear geometry and choice of 980-nm dichroics, as a first alignment step we can conveniently first couple the 795-nm and 1475-nm laser light to the 780-nm and 1530-nm fibers, respectively. We then introduce a 780-nm laser, copropagating with the pump lasers, to generate a stimulated 1530-nm beam that is detectable on an amplified InGaAs photodiode. After walking the 780-nm laser beam to maximize the stimulated emission, we couple the 780-nm laser and stimulated emission to their respective fibers. This ensures that the spatial modes we collect meet the phase-matching condition, and are thus correlated. Finally, we remove the 780-nm laser and maximize the spontaneous bi-photon rate, detecting on SNSPDs.

Our general pump regimes are defined by the intermediate level detuning $\Delta_{1}$ set by the 795-nm pump frequency (referenced to $|5S_{1/2},F=2\rangle\rightarrow|5P_{1/2},F'=1\rangle$). In each regime we sweep the two-photon pump detuning $\Delta_{2}$ (referenced to $|5S_{1/2},F=2\rangle\rightarrow|5P_{1/2}\rangle\rightarrow |4D_{3/2},F''=3\rangle$) across the atomic resonance to address different hyperfine levels in the excited states and different velocity groups. Characterization and optimization of performance metrics including the spectrum of the 1530-nm photons, photon pair coincidence rate, cross-correlation and heralded auto-correlation are discussed in the Supplementary Information Section 1.

\subsection*{Solid-state quantum memory}
\vspace{-5pt}
The solid-state quantum memory at node B is based on a bulk $^{166}$Er$^{3+}$:YVO$_{4}$ crystal (Gamdan Optics) cooled to 12~mK base temperature in a dilution refrigerator. The choice of YVO$_{4}$ among other host crystals (Table~\ref{tab:hostcrystalsummary}) is for the closest spectral matching to the Rb telecom emission. The magnetic field is provided by a vector magnet (AMI 1-0.4-0.4T) mounted in the same dilution fridge. The crystal is mounted on top of a gold coated mirror (Thorlabs NB05-L01) with its $c$-axis perpendicular to the incident light and parallel to the external magnetic field. The light is collimated from an SMF-28 single mode fiber and focused onto the crystal-mirror interface with a beam waist of 3.5~$\mathrm{\mu}$m. The reflected light is collected through the same fiber (Fig.~\ref{fig:setup}c). The optics are assembled on a custom-made mounting system, including a nanopositioner for in-situ alignment at cryogenic temperatures.

With an isotopically purified $^{166}$Er$^{3+}$ concentration of 15~ppm and an effective crystal path length of 8~mm, the crystal measures an optical depth of 4.5 at 1530~nm. The laser used to address the Er$^{3+}$ transition is an external cavity diode laser (Toptica DL pro), frequency-stabilized to a ULE reference cavity (Stable Laser Systems) via the Pound–Drever–Hall technique. We use two fiber-coupled acousto-optical modulators (AOM) in series for laser amplitude and frequency modulation, each with a center RF frequency of 200~MHz. A polarization controller is used to optimize the light polarization incident onto the crystal.

For optical AFC preparation, we sweep the optical pumping laser discretely with a programmable periodicity $\Delta_{\mathrm{AFC}}$ over a 100-MHz spectral bandwidth. At each frequency, the pump is turned on for 64$\mathrm{~\mu}$s. We start from the frequencies at the center of the transition ($f_{-1}=f_{0}-\Delta_{\mathrm{AFC}}$, $f_{0}$ and $f_{+1}=f_{0}+\Delta_{\mathrm{AFC}}$), then jump back and forth between negatively and positively detuned frequencies. Spectral hole-burning depletes spin population at each pump frequency $f_{n}$. Adjacent pumps at frequencies $f_{n-1}$ and $f_{n+1}$ only create new spectral holes without affecting the spectral hole at $f_{n}$ (more details see Supplementary Information Section 2). The entire holeburning procedure is repeated for 600 times. Following a wait time of 500~ms, the memory is ready for storing photons. After the storage duration of a few seconds, there is a $\sim$10~ms rest time to allow the crystal to fully thermalize before the next AFC preparation cycle starts (Fig.~\ref{fig:control}).

As a prerequisite for AFC preparation, we perform spectral hole-burning measurements and develop a theoretical model to explain the results, by simulating the energy level structure (i.e. super-hyperfine levels) in $^{166}$Er$^{3+}$:YVO$_{4}$. With the insight, we optimize AFC preparation using the storage efficiency for a weak coherent pulse as the main metric and further investigate the efficiency dependence on the AFC bandwidth. The details of these experiment and simulation results are provided in the Supplementary Information Section 2.

\subsection*{Source--Memory temporal gating}
\vspace{-5pt}
To maintain a low accidental coincidence (background noise) when measuring the retrieved telecom photons, we implement a temporal gating scheme for both the 780-nm heralding channel and the 1530-nm signal channel. In our gating scheme, the 1530-nm photons are gated optically using an 80-MHz AOM (Aerodiode), while the 780-nm heralding photons are gated electronically using the Time Tagger. The two channels share an identical gating pattern---over a 2-$\mu$s cycle (twice the storage time), the gate is opened for 0.8~$\mu$s and closed for 1.2~$\mu$s. This sequence repeats throughout the photon storage experiment. With this gating, the accidental coincidence around the photon echo detection window is significantly suppressed, which recover the signal-to-noise ratio as well as the non-classical cross-correlation between the 780-nm heralding and the 1530-nm echo photons. Additional details on this gating scheme, including a model for the coincidence background, a comparison of coincidence histograms with and without gating, and an analysis on the remaining noise contributions, are provided in the Supplementary Information Section 3.

\subsection*{Source--Memory efficiency}
\vspace{-5pt}
The overall system storage efficiency is defined as the ratio between the echo coincidence counts and the source coincidence counts. For both quantities, we use the integrated coincidence counts after subtracting the fitted background. The system efficiency is affected by several factors. First, the duty cycle of the photon gating is 19$\%$. Second, the combined loss of optical components (listed in Table~\ref{tab:lossbudget}) leads to an efficiency of 5.6$\%$. Removing these two contributions yields the internal storage efficiency, which reflects the intrinsic telecom photonic interfaces of  the source and the memory. As shown in Fig.~\ref{fig:efficiency} for various two-photon pump detunings $\Delta_{2}$ at the source, the variation of the internal efficiency can be understood from three factors:

\begin{enumerate}[label=(\roman*)]
\item AFC memory efficiency. For a Gaussian-shaped input photon with a 50-MHz FWHM, the AFC memory exhibits a storage efficiency of 5$\%$. Based on our measurements with a 43-MHz-FWHM weak pulse, this estimate is reasonable.
\item Polarization selection. The photon-pair source produces randomly polarized photons, whereas the memory absorption is maximized for a fixed polarization that is aligned to the Er transition dipole moment. This selection reduces the effective absorption by 50$\%$.
\item Spectral selection (internal filtering). About 20$\%$ of the 1530-nm photon spectrum falls within the 100-MHz bandwidth of the AFC memory. This fraction varies with the two-photon pump detuning $\Delta_{2}$, as it modifies the photon spectrum. 
\end{enumerate}

Combining all three factors, at the optimal two-photon pump detuning (i.e. $+$703~MHz), the measured 0.005$\%$ system efficiency and  0.53$\%$ internal efficiency agree well with the estimation above.

\section*{Acknowledgments}
\vspace{-5pt}
We thank Ian Chin, Shankar G Menon, Noah Glachman, Shobhit Gupta, Jackson Swartz, and Kevin Singh for facilitating experiments at different stages. We thank Reet Mhaske for modeling the memory crystal and Allen Zang for fruitful discussions. We gratefully acknowledge funding from the NSF QLCI for Hybrid Quantum Architectures and Networks (NSF award 2016136). H.B. acknowledges funding by the NSF Quantum Interconnects Challenge for Transformational Advances in Quantum Systems (NSF award 2138068), and the NSF Career program (NSF award 2238860). T.Z. acknowledges funding by the NSF Career program (grant number 1944715) and Army Research Office (ARO) grant W911NF2010296.

\section*{Author contributions}
\vspace{-5pt}
Y.C., D.G., N.P.T., A.K. built the setup and performed the experiments. Y.C., D.G., N.P.T. analyzed the data. All authors discussed and interpreted the results. All work was supervised by H.B. and T.Z.. Y.C. prepared the initial manuscript and all authors contributed to the final manuscript.

\section*{Data availability}
\vspace{-5pt}
All data related to this study are available from the corresponding author upon request.

\section*{Code availability}
\vspace{-5pt}
All analysis codes related to this study are available from the corresponding author upon request.

\onecolumngrid
\renewcommand{\tocname}{Supplementary Information}
\renewcommand{\appendixname}{Supplement}
\renewcommand{\figurename}{Fig.}
\renewcommand{\tablename}{Table}
\setcounter{equation}{0}
\setcounter{figure}{0}
\setcounter{table}{0}
\renewcommand{\theequation}{S\arabic{equation}}
\renewcommand{\thefigure}{S\arabic{figure}}
\renewcommand{\thesubsection}{S\arabic{subsection}}
\renewcommand{\theHfigure}{SI\arabic{figure}}
\renewcommand{\thetable}{S\arabic{table}}
\renewcommand{\theHtable}{SI\arabic{table}}
\clearpage
\begin{appendices}

\clearpage

\section*{Supplementary Information}

\begin{figure*}[h!]
\centering
\includegraphics{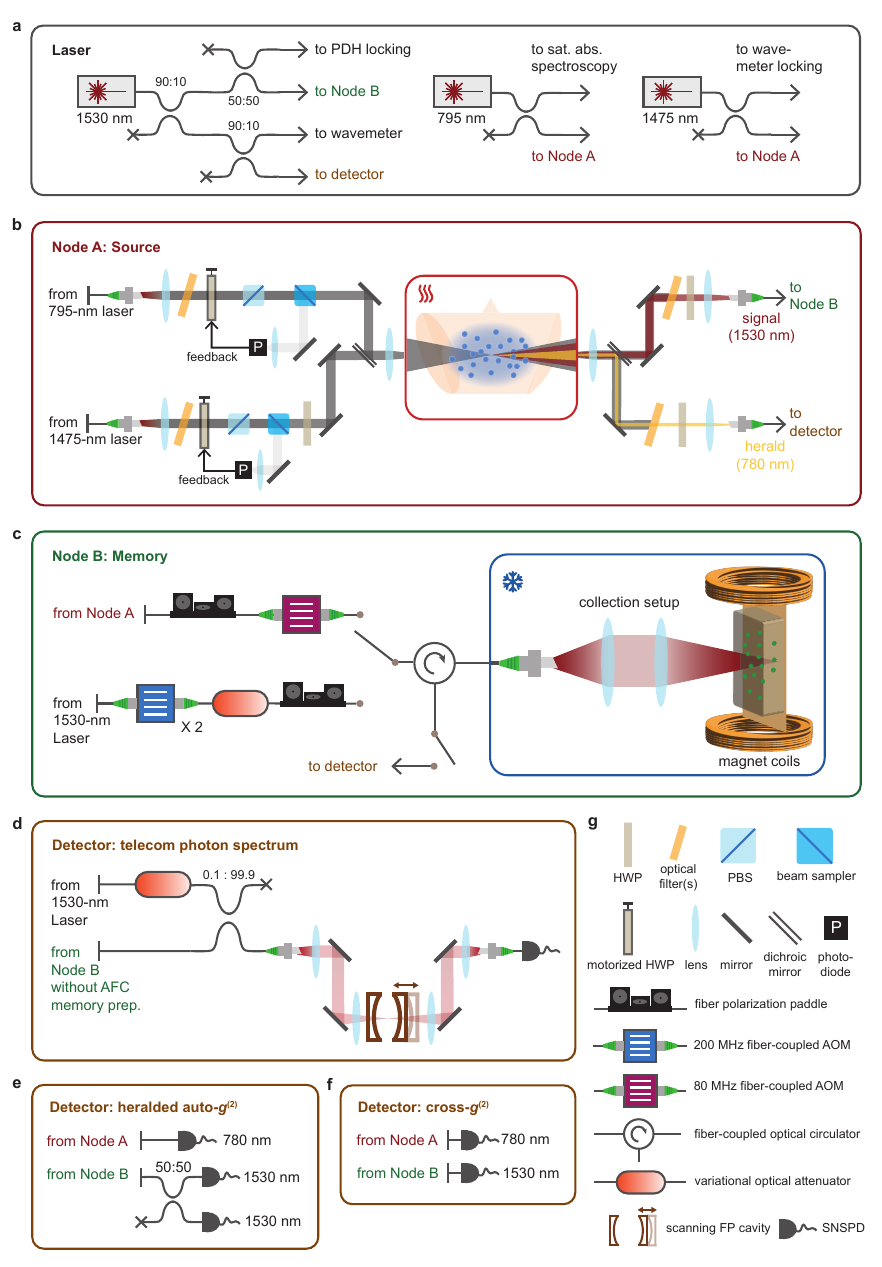}
\caption{\textbf{Experimental setup.} \textbf{a}, Laser module. \textbf{b}, \textbf{c}, Two quantum nodes are located in two laboratories separated by 35 m. \textbf{d}, \textbf{e}, \textbf{f}, Three detector modules for different measurements, all located in the same lab as node A. \textbf{g}, Main devices as legends.}
\label{fig:setup}
\end{figure*}

\begin{figure*}[h!]
\centering
\includegraphics{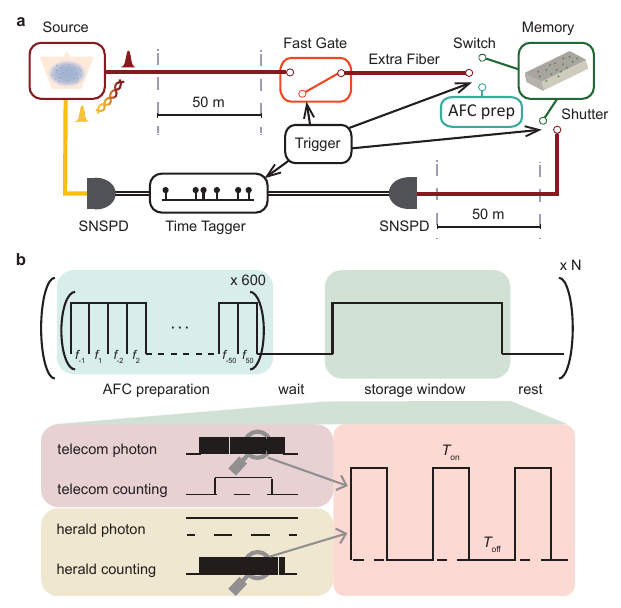}
\caption{\textbf{Experimental control and sequence.} \textbf{a}, Layout overview, details of the source and memory module are shown in Fig.~\ref{fig:setup}. Fast Gate: physical gating via an AOM for the 1530-nm channel and electronic gating at the Time Tagger for the 780-nm channel realize temporal gating for minimizing echo background. Switch: toggles between the AFC preparation mode and the storage acceptance mode. Trigger: all trigger signals are sent from the arbitrary waveform generator (HDAWG) and thus synchronized with the complete pulse sequence. The SNSPD and Time Tagger are located in the same lab as the source in node A; while the gating AOM, Switch and HDAWG are located in the same lab as the memory in node B. \textbf{b}, General pulse sequence of the networking experiment. The sequence begins with AFC preparation, followed by a waiting period. The storage acceptance window then opens, allowing the memory to receive 1530-nm signal photons from the source. A rest period is included at the end as a buffer. Within the storage window, a synchronized sequence divides it into alternating on-off blocks for implementing the temporal gating scheme.}
\label{fig:control}
\end{figure*}

\subsection{Photon source characterization and optimization}

Photon sources based on atomic vapors have a vast parameter space, with many viable operating regimes that each have unique optical properties. For example, sources based on electromagnetically-induced transparency (EIT) provide on-demand sub-MHz bandwidth photons~\cite{2005_Eisaman_EIT}, while spontaneous four-wave mixing (4WM) -based sources can offer polarization-entangled photon pairs with GHz bandwidth~\cite{2011_Rolston_polarization, 2021_Davidson_4WM}. Easily accessible control knobs---vapor cell temperature, laser detuning and intensity, polarization, beam angle---allow for relatively simple switching to different regimes. Thus the atomic vapor photon source is highly customizable, with the user choosing the regime best tailored to their application. 
However, tuning these experimental parameters can have complex, confounding effects on pertinent photon source metrics---rate, heralding efficiency, single photon purity, spectrum---that are still not well understood, and largely unexplored. Here we explore 4WM in $^{87}$Rb using the level structure in Fig.~\ref{fig:architecture}b, with a special focus on optical frequency and bandwidth to maximize overlap with our quantum memory.

\subsubsection{Photon spectrum characterization} 

As shown in Fig.~\ref{fig:setup}d, to measure the spectrum of the 1530-nm photons, we use a Fabry-P\'erot interferometer with scanning piezos (Thorlabs SA 30-144) and detect on an SNSPD. The interferometer has a nominal spectral resolution < 1~MHz and a measured free spectral range (FSR) of 2.80(4)~GHz. The piezo drifts by up to 25~MHz every half hour; to avoid broadening of our measured spectra, a frequency-locked reference laser at 1530~nm is combined with the signal to provide a stable frequency reference during data-taking. The spectra were taken in shorter batches and combined together in post-processing using the laser as an offset reference.

The cavity was aligned following the manual, with the added step of extinguishing the TEM01 mode (confirmed with a camera image of the outgoing beam) to optimize coupling into the fiber going to the SNSPD. The cavity piezo voltage is controlled with the accompanying driver (Thorlabs SA201B) using a triangle waveform with a 10 ms rise time and a 30 V span. The setup has an overall 18\% efficiency between the fiber outcoupling and the SNSPD. In future work the nonlinearity of the piezo scan may be calibrated to yield a more accurate absolute frequency and spectrum shape.

\begin{figure*}[h!]
\centering
\includegraphics{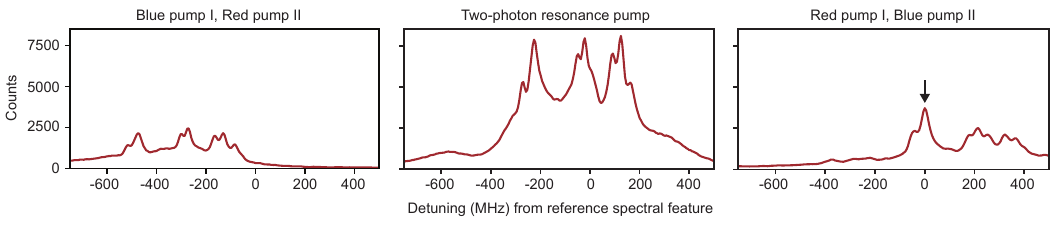}
\caption{\textbf{Source 1530-nm spectra with different pump regimes.} From left to right are the spectra for Blue pump-I, Red pump-II ($\Delta_{1}>0, \Delta_{2}<0$); Two-photon resonance pump ($\Delta_{1}=\Delta_{2}=0$); Red pump-I, Blue pump-II ($\Delta_{1}<0, \Delta_{2}>0$). All three spectra are frequency-referenced to the target spectral feature (indicated here by the black arrow) used for source--memory matching and networking in the main text.}
\label{fig:source_spectra}
\end{figure*}

\begin{figure*}[h!]
\centering
\includegraphics{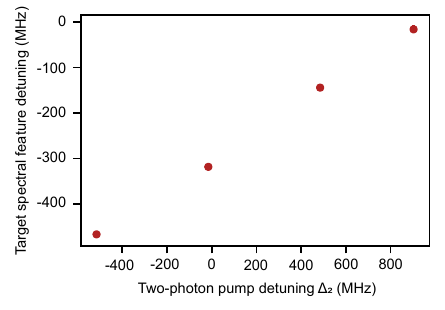}
\caption{\textbf{Source 1530-nm photon target spectral feature frequency with different two-photon pump detunings.} Error bars are smaller than the points. Scanning across the two-photon pump detuning ($\Delta_2$)  with fixed $\Delta_1=-817\,$~MHz, the frequency of output 1530-nm photons accordingly shifts linearly. Note that the slope is less than one, indicating that the overall shift is split between the 1530-nm and 780-nm photons, for conservation of total momentum and energy.}
\label{fig:source_freq_sweep}
\end{figure*}

As seen in Fig.~\ref{fig:source_spectra}, the telecom photon spectrum changes in both shape and amplitude based on the pump laser detunings. Three overall manifolds are visible and correspond to transitions from $|4D_{3/2}\rangle$ to $|5P_{3/2}, F'={3,2,1}\rangle$. The change in relative heights of the main features across different pump regimes corresponds to the change in relative Rabi frequencies of different velocity classes; more investigation is necessary to quantify this relationship. We choose $\Delta_1<0$, $\Delta_2>0$, as a blue two-photon detuning generates higher-frequency telecom photons (Fig.~\ref{fig:source_freq_sweep}), allowing for spectral matching with the crystal at a lower applied magnetic field. Additionally, more of the light is concentrated in one peak (approximately 20-25$\%$ of the entire spectrum) with a bandwidth compatible with the memory (100~MHz), allowing for a higher rate and heralding efficiency after storage and retrieval.

\subsubsection{Laser frequency dependence of rate and cross-correlation}

\begin{figure*}[h!]
\centering
\includegraphics{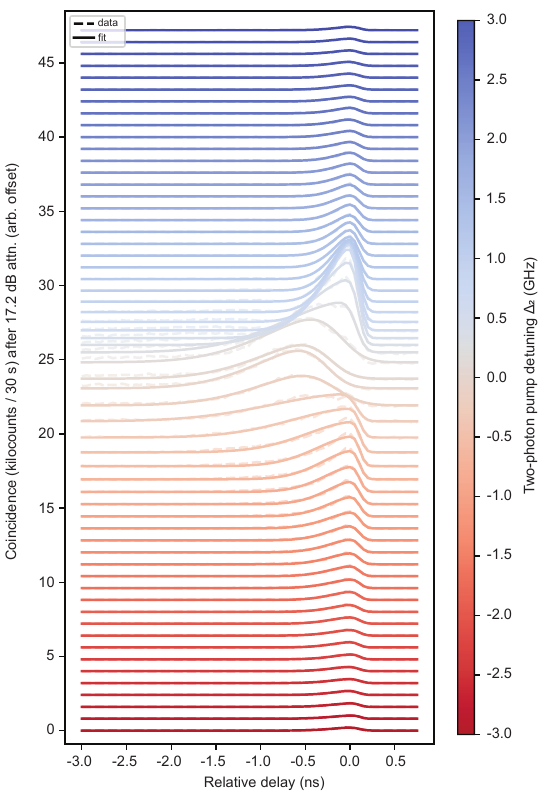}
\caption{\textbf{Source coincidence histograms.} Coincidences between the 780-nm heralding and the 1530-nm signal channel, with 50-ps bin and 30-second integration time. The two-photon pump detuning $\Delta_{2}$ is swept by varying the 1475-nm laser piezo voltage, while the 795-nm laser is constant at $\Delta_{1}=-$817~MHz detuned from the $|5S_{1/2},F=2\rangle \rightarrow |5S_{1/2},F=1\rangle$ transition.}
\label{fig:crossg2_data}
\end{figure*}

\begin{figure*}[h!]
\centering
\includegraphics{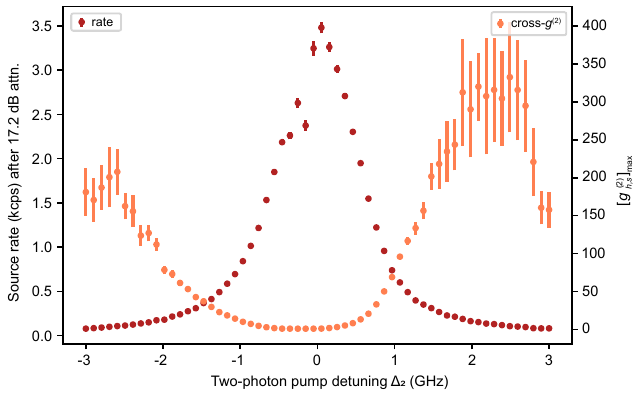}
\caption{\textbf{Source coincidence rate and cross-correlation $g^{(2)}$ of photon pairs before storage, as function of the two-photon pump detuning $\Delta_{2}$.} Values and errorbars are extracted from fits in Fig.~\ref{fig:crossg2_data}. The rates are measured after 17.2~dB of optical attenuation, to avoid SNSPD saturation and data overflow in the photon counter's internal buffer.}
\label{fig:source_tradeoff_full}
\end{figure*}

We sweep the 1475-nm laser to optimize the second-order cross-correlation $[g_{h,s}^{(2)}]_{\mathrm{max}}$ and coincidence rate of the source (Fig.~\ref{fig:crossg2_data} and Fig.~\ref{fig:source_tradeoff_full}). Within the Doppler-broadened linewidth ($\sim$~1~GHz), we see high rates at the cost of low cross-$g^{(2)}$, due to resonant excitation and the resulting scattering of uncorrelated photons. Outside of the Doppler linewidth, the desired 4WM process dominates. As detuning is increased, the variation in Rabi frequencies becomes more uniform across different velocity classes ($\Omega \propto 1/\Delta$), causing more atoms to participate in the collectively-enhanced 4WM, both improving the cross-$g^{(2)}$ and narrowing the correlation time~\cite{2021_Davidson_4WM}. However, because changing the detuning has multiple confounding effects (on average Rabi frequency, distribution of Rabi frequencies, phase-matching conditions, collective enhancement), more investigation is necessary (e.g. a multidimensional sweep with both detuning and optical power) to isolate these effects.

We note an asymmetry in rate and cross-$g^{(2)}$ between red and blue two-photon detunings. In the case of a red-detuned pump I ($\Delta_1<0$), blue detuning from the two-photon resonance ($\Delta_2>0$) outperforms red detuning from the two-photon resonance. We also observe the opposite behavior for blue-detuned pump I ($\Delta_1>0$); in this case $\Delta_2 < 0$ performs better. This asymmetry between positive and negative $\Delta_2$ may be explained by the increased range of velocity classes with comparable two-photon Rabi frequencies when both positive and negative detunings are present, but further investigation is necessary. We choose the former case ($\Delta_1<0, \Delta_2>0$), as a positive overall detuning yields photons frequency-matched with the memory at a lower applied magnetic field.

The non-classicality of the photon pair can be benchmarked by the Cauchy-Schwarz inequality~\cite{1986_Walls_CauchySchwarz}: non-classical correlations means violation of 

\begin{equation}
\label{eq:CSinequality}
\mathcal{R} = \frac{\left(g^{(2)}_{h,s}\right)^{2}}{g^{(2)}_{h,h}g^{(2)}_{s,s}}\leq 1
\end{equation}
where $g^{(2)}_{h,h}$ and $g^{(2)}_{s,s}$ are the unheralded auto-correlation of the heralding photon and the signal photon respectively. We use the thermal photon statistics $g^{(2)}_{Th, Th}=2$ as an upper bound to get a classical threshold of $g^{(2)}_{h,s}\leq 2$. The regime we choose ($\Delta_{1}=-$817~MHz, $\Delta_{2}=+$903~MHz) in the main text violates the Cauchy-Schwarz inequality by more than three orders of magnitude ($\mathcal{R}\approx4\times10^{3}$). We later use the same threshold to characterize the echo coincidence after storage and retrieval from the memory.

\subsubsection{Laser power dependence of rate and cross-correlation}

\begin{figure*}[h!]
\centering
\includegraphics{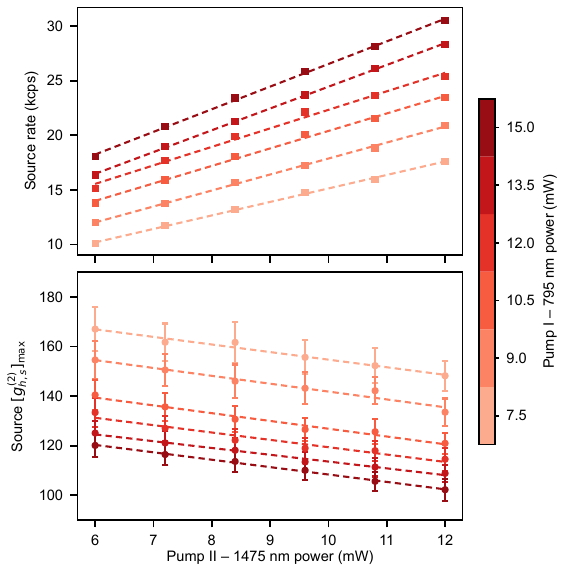}
\caption{\textbf{Source pump power sweeps.} Dashed lines are linear fits to the data.}
\label{fig:source_powersweep}
\end{figure*}

In Fig.~\ref{fig:source_powersweep}, we demonstrate linear scaling of coincidence rates as a function of both laser powers, while still maintaining highly non-classical cross-$g^{(2)}$. For the pump regime we operate in with maximum power (Fig.~\ref{fig:architecture}c), we can estimate mean photon pair number based on the detection rate for the heralding channel and the signal channel (423~kcps and 2333~kcps respectively),  the bi-photon detection rate (46~kcps) and the correlation time (0.32(2)~ns): $\langle n \rangle =$~423~kcps$\times$2333~kcps$/$46~kcps$\times$0.32~ns = 0.007 $\ll$ 1. This indicates that future work could see enhanced rates with boosted laser powers while staying in the low pump regime. See Fig.~\ref{fig:autog2}c, d for corresponding values of the auto-$g^{(2)}(0)$ as functions of laser powers.

\subsubsection{Heralded auto-correlation characterization}

To characterize the single-photon nature of the signal photons, we perform a heralded Hanbury-Brown-Twiss (HBT) measurement (Fig.~\ref{fig:network}a), in which the signal photon channel is split with a fiber beam splitter, and coincidences between the two output channels are counted, with conditioning by the heralding channel on both signal channels. Coincidences corresponding to $\Delta n = 0$ occur when both signal channels receive a click within a time window $\Delta t$ after a click on the herald channel. Coincidences corresponding to $\Delta n \neq 0$ occur when both signal channels receive a click after two different heralding photons, where the signal clicks are in the windows $\Delta t$ after their respective heralds, and the two heralds are spaced by $\Delta n$ clicks on the heralding channel.

\begin{figure*}[h!]
\centering
\includegraphics{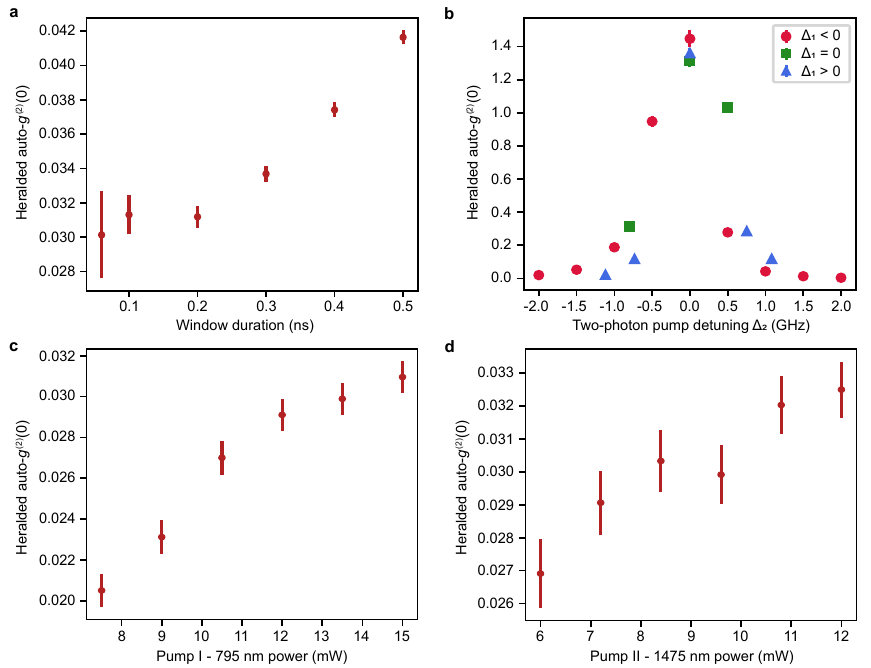}
\caption{\textbf{Heralded auto-$g^{(2)}(0)$ characterization.} \textbf{a}, Dependence on acceptance window duration $\Delta t$. \textbf{b}, Dependence on pump laser detunings. Red circles: $\Delta_1=-$817~MHz; green squares: $\Delta_1=$~0; blue triangles: $\Delta_1=+$705~MHz. \textbf{c}, Dependence on pump I power. \textbf{d}, Dependence on pump II power.}
\label{fig:autog2}
\end{figure*}

To choose the acceptance window duration $\Delta t$ for Fig.~\ref{fig:network}a, we first found the heralded auto-$g^{(2)}(0)$ as a function of $\Delta t$ in post-processing (Fig.~\ref{fig:autog2}a). We find that the photons maintain single photon behavior ($g^{(2)}(0) < 0.5$) across the correlation time ($\approx$~0.32~ns), and choose $\Delta t=$~0.2~ns for the rest of the analysis.

We further characterize heralded auto-$g^{(2)}(0)$ as a function of various experimental parameters. When measuring heralded auto-$g^{(2)}(0)$ as a function of laser frequency (Fig.~\ref{fig:autog2}b), the trend closely resembles the rate (Fig.~\ref{fig:source_tradeoff_full}) as expected. Here, only when the pumps are outside the Doppler-broadened linewidth do generated signal photons show single-photon behavior ($g^{(2)}(0) <$~0.5). 
In our detuned regime, we also find that the photons demonstrate strong single-photon behavior across our entire range of available laser powers (Fig.~\ref{fig:autog2}c, d), indicating that with boosted laser power, future work could see enhanced rates while still maintaining single-photon behavior.

\subsection{Quantum memory characterization and optimization}

\subsubsection{Spectral hole-burning}
We experimentally examine the shape and lifetime of a spectral hole by applying the standard pump-wait-probe sequence. To better mimic the actual AFC preparation sequence, the pump pulses are run in a repetitive manner with off times between the pulses. We implemented the same AFC preparation sequence in Fig.~\ref{fig:control}, with the pump laser turned off for all but one frequency (at which we characterize spectral hole-burning). Effectively, we have a 64 $\mu$s optical pump at a single frequency, followed by an off time of 6.3 ms. After repeating this pump cycle 600 times, we wait for a tunable amount of time ($T_{\mathrm{wait}}$) then probe the spectral hole for 768~$\mu$s. We notice that the hole spectrum only stabilizes after a few minutes of repeatedly running the pump-wait-probe sequence, indicating certain spin dynamics occurring on the minute time scale.

\begin{figure*}[h!]
\centering
\includegraphics{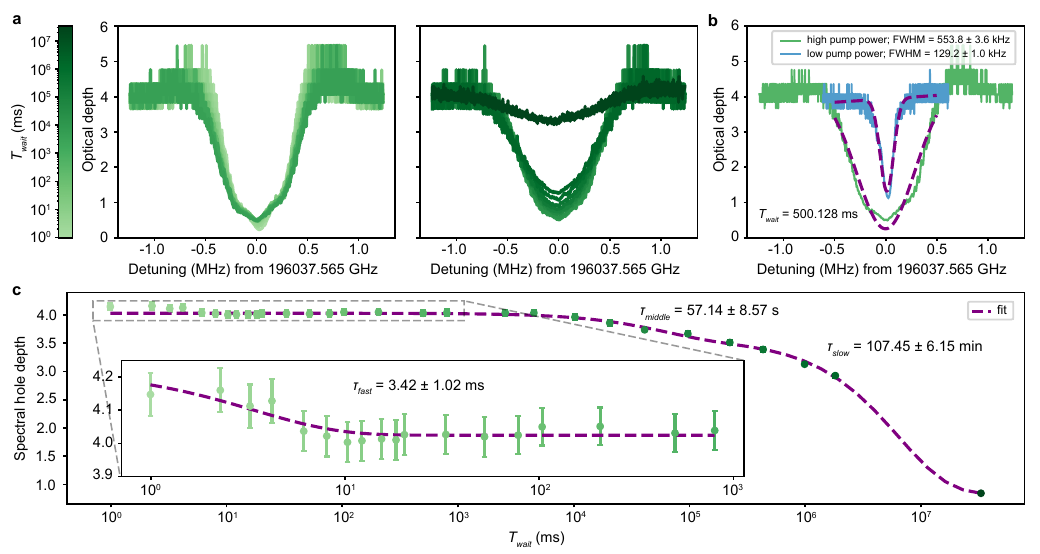}
\caption{\textbf{Spectral hole-burning in $^{166}$Er$^{3+}$:YVO$_{4}$.} \textbf{a}, Spectral hole shape (with high pump power) with changing wait time ($T_{\mathrm{wait}}$) which is indicated by different colors. The left sub-plot shows $T_{\mathrm{wait}}$ from 0.99~ms to 4.528~s while the right plot shows $T_{\mathrm{wait}}$ from 10.241~s to 9.16~hours. \textbf{b}, Comparison between spectral holes with high/low pump power, $T_{\mathrm{wait}}$ is set to be 500~ms. \textbf{c}, Spectral hole depth decays over $T_{\mathrm{wait}}$. There are three exponential decay components. The fast decay is shown in the inset. The other two are shown in the main plot.} 
\label{fig:holeburning}
\end{figure*}

Figure~\ref{fig:holeburning}\textbf{a} displays the shape of the spectral hole at varying wait times after the pump. There are two bumps at  both red and blue detunings of the main hole, which reassemble two groups of anti-holes. The detuning of both groups of anti-holes indicates that the hole-burning involves a double-$\Lambda$ system, where the difference between the ground state splitting and that of the excited state is sub-MHz, and is on the order of a few 100s~kHz. In Fig.~\ref{fig:holeburning}b, we show the spectral hole shapes under two different pump powers. The Rabi frequency of the high power pump field (Fig.~\ref{fig:holeburning}a) is estimated to be $\Omega_{h}\sim 2\pi\times$1~MHz, while that of the low power pump is ten times smaller thus $\Omega_{h}\sim 2\pi\times$100~kHz. Pumping with a low power close to that used in the probe sequence gives a shallower and narrower spectral hole. The fitted linewidth of 129(1)~kHz implies an optical coherence time up to 5~$\mu$s~\cite{2008_Geneva_holeburning} and sets a lower limit on the AFC teeth spacing. The hole width is broadened to 553(4)~kHz under high pump power. This power-broadening of the hole provides a means to optimize the AFC finesse and the storage efficiency, which will be discussed in sub-section S2.3.

Figure~\ref{fig:holeburning}c shows the spectral hole relaxation dynamics. We plot the hole depth as a function of the wait time. The inset zooms into a shorter time scale within 100 ms and fits to an exponential decay with $\tau_{\mathrm{fast}}=$~3.42~$\pm$~1.02~ms, which is indicative of the optical lifetime of the excited state. Beyond this fast optical decay, there are two exponential decays with $\tau_{\mathrm{middle}}=$~57.15~$\pm$~8.57~s and $\tau_{\mathrm{slow}}=$~107.45~$\pm$~6.15~min and a weight percentage of 14.5$\%$ and 85.5$\%$, respectively. The fitting further gives a constant hole depth background, which suggests an even longer decay, on the order of days. Such a long hole lifetime is only possible with nuclear spins in the host matrix, and is unlikely to be from the $^{166}$Er$^{3+}$ electron spins. Under the operating magnetic field of 1 T, the ground state $^{166}$Er$^{3+}$ electron spin splitting is 46.7~GHz. With the estimated effective temperature of the crystal as 150~mK, the Er spins are 99.99997$\%$ polarized, therefore, they cannot account for the long hole lifetimes we observed. 

Taken together, the results in Fig.~\ref{fig:holeburning}a and c indicate that the efficient hole-burning in our experiment takes place via long-lived nuclear spin states, with the optical transitions of the associated nuclear spin levels spaced by sub-MHz. Such narrow and closely packed spectral hole features make it possible to create a sharp-teeth and broad bandwidth AFC up to the inhomegeneous linewidth of the Er telecom transition. 


\subsubsection{$^{166}$Er$^{3+}$:YVO$_{4}$ energy levels for spectral hole-burning}

We study the detailed energy levels of the $^{166}$Er$^{3+}$:YVO$_{4}$ crystal, including the super-hyperfine levels to explain the efficient spectral hole-burning we have achieved in this work.

\begin{figure*}[h!]
\centering
\includegraphics{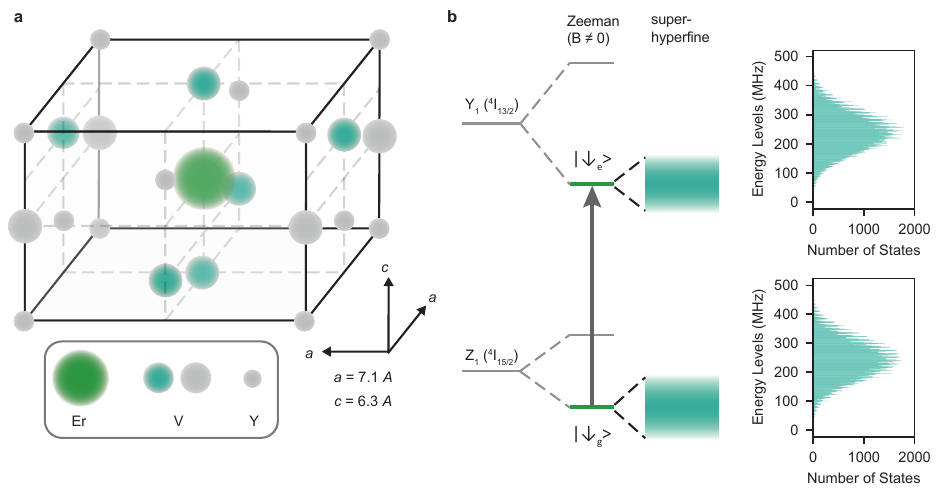}
\caption{\textbf{$^{166}$Er$^{3+}$:YVO$_{4}$ energy levels.} \textbf{a}, The crystal lattice structure of YVO$_{4}$ where $^{166}$Er$^{3+}$ substitutes for yttrium in a single site with non-polar ($D_{2d}$) symmetry. Around $^{166}$Er$^{3+}$, there are two nearest neighbor and four next-nearest neighbor vanadium (V$^{5+}$) ions. \textbf{b}, Simulated Er$^{3+}$-V$^{5+}$ super-hyperfine energy level distribution for the lower Zeeman electron spin states in the optical ground and excited states ($|\downarrow_{g}\rangle$ and $|\downarrow_{e}\rangle$). The frequency bin in both energy band plots is 1~MHz. The optical transitions between the two bands show frequency differences well below 1~MHz.}
\label{fig:crystallevel}
\end{figure*}

In an Er:YVO$_{4}$ crystal, $^{166}$Er$^{3+}$ substitutes for yttrium in a single site. Each Er$^{3+}$ ion experiences a nuclear spin environment consisting of 99.8$\%$ $^{51}$V and 100$\%$ $^{89}$Y isotopes with nuclear spins of 7/2 and 1/2, respectively. The interactions of erbium electron spin with yttrium and vanadium nuclear spins result in a splitting of each Er Zeeman level into numerous super-hyperfine levels. We model the super-hyperfine interactions using the Hamiltonian:

\begin{equation}
\label{eq:shhamiltonian}
\begin{aligned}
H=\mu_{B}\vec{B}\cdot&\bar{\bar{g}}^{Er}\cdot\vec{S}^{Er}   + \sum_i\mu_{N}\vec{B}\cdot\bar{\bar{g}}^{i}\cdot\vec{I^{i}}  + \sum_i Q_i (\vec{I^{i}})^{2}  \\ & -\sum_i \frac{\mu_0\mu_B\mu_N}{4\pi}\Big(3 (\vec{r}_i\cdot \bar{\bar{g}}^{Er}\cdot \vec{S}^{Er}) (\vec{r}_i\cdot \bar{\bar{g}}^{i}\cdot \vec{I}^{i}) \frac{1}{r_i^5}- (\bar{\bar{g}}^{Er}\cdot \vec{S}^{Er})(\bar{\bar{g}}^{i}\cdot \vec{I}^{i})\frac{1}{r_i^3} \Big)
\end{aligned}
\end{equation}
where the summation $i$ runs over the nuclear spins $\vec{I}^{i}$ at position $\vec{r}_{i}$ relative to the Er$^{3+}$ ion, with $\bar{\bar{g}}^{i}$ and $Q_{i}$ as the $g$-factor and the quadruple coupling strength of the nuclear spin. The eigenstates give rise to manifolds of super-hyperfine levels within each Zeeman branch of the Er$^{3+}$ electron spins.

We analyze two super-hyperfine couplings: Er-V and Er-Y, and consider the interactions up to the second nearest neighbors. For Er-V coupling, the Er$^{3+}$ has two nearest neighboring V$^{5+}$ ions at a distance 3.1~\r{A} along the crystal symmetry $c$-axis, followed by four next-nearest neighbors at a distance of $\approx$~3.9~\r{A}. Our calculation of the above Hamiltonian is restricted to these six V nuclear spins. The Er$^{3+}$ electron spin is assumed to be polarized along the axis of the applied magnetic field, which, in our case, is parallel to the $c$-axis. The total magnetic field on each V$^{5+}$ is a vector sum of the applied magnetic field and the small field produced by Er$^{3+}$. Using the nuclear $g_{V}$~=~1.6 for Vs along the symmetry axis and the $Q=$~171~kHz and 165~kHz for the nearest and the next-nearest Vs respectively~\cite{2024_Ruskuc_Vspin}, we approximate the super-hyperfine energy level spectra for the lower Zeeman branches in the Z$_{1}$ ($|\downarrow_{g}\rangle$) and Y$_{1}$ ($|\downarrow_{e}\rangle$) levels. As shown in Fig.~\ref{fig:crystallevel}, the calculation gives a band of energy levels with a total width of roughly 500~MHz. The center of the band has the highest density of the energy levels. The spacings between neighboring levels (with a single nuclear spin flip, i.e. $\Delta m_{V}=\pm 1$ for one out of the six Vanadium) are in the range of 10-12~MHz. Specifically, the spacing between the $|-1/2\rangle_{g}$ and $|+1/2\rangle_{g}$ V nuclear spin levels is 10.9~MHz. Next, we calculate the spacings of optical transitions between neighboring V nuclear spin levels (i.e. the difference between the spin level spacings in the ground and excited states) for the two nearest Vs as 351.3~kHz, and for the four next-nearest Vs as 48.4~kHz. Given a $\sim$~1~MHz optical Rabi frequency of the pump laser, optical transitions within $\sim$~0.5~MHz vicinity of the pump are excited, transferring populations to nuclear spin levels further detuned from the spin level addressed by the pump laser. This hole-burning mechanism would result in anti-hole features appear at detunings that are integer multiples of the spacings, for instance 702~kHz---twice the spacing of 351.3~kHz and larger than the Rabi frequency. The measured spectral hole shape in Fig.~\ref{fig:holeburning} agrees well with this explanation, showing anti-hole features at $\sim$~600-700~kHz detuning from the spectral hole.

The same calculation for the Er-Y ($g_{Y}=-$0.137) interaction yields spacings of the neighboring optical transitions of 4.4~kHz for the nearest Ys and 0.4~kHz for the next-nearest Ys. Such spacings are too small to allow for spectral hole-burning as the pump laser would simultaneously excite all Y nuclear spin levels. Based on these energy level calculations, we conclude that the spectral holes observed in our $^{166}$Er$^{3+}$:YVO$_{4}$ crystal are primarily contributed by the nearest Vanadium nuclear spins. The full population transfer dynamics in the hole-burning process remains a subject for future studies.

\subsubsection{AFC storage characterization}
We characterize the performance of the AFC quantum memory with a weak coherent pulse as the input. The input pulse has a Gaussian shape modulated by the same AOMs used for spectral hole-burning and AFC preparation. An additional variational optical attenuator (VOA) is used to further attenuate the intensity of the input to make sure the mean photon number is at single-photon level (Fig.~\ref{fig:setup}). We choose a wait time of 500.608~ms between the AFC preparation pulses and the photon storage. 

The theoretical AFC storage efficiency for a comb with Gaussian-shaped teeth is given by~\cite{2009_Gisin_AFC_OGTh}:

\begin{equation}
\label{eq:AFCefficiency}
\eta_{\mathrm{AFC}} = \left(\frac{d}{F}\right)^{2}\mathrm{exp}\left(-\frac{d}{F}\right)\mathrm{exp}\left(-\frac{\pi^{2}/(2\mathrm{ln}2)}{F^{2}}\right) e^{-d_{0}}
\end{equation}
where $d$ is the optical depth of the teeth, $d_{0}$ is the background optical depth and $F$ is the finesse of the comb. From the hole-burning measurement, we have $d$ = 4.5.
\begin{figure*}[h!]
\centering
\includegraphics{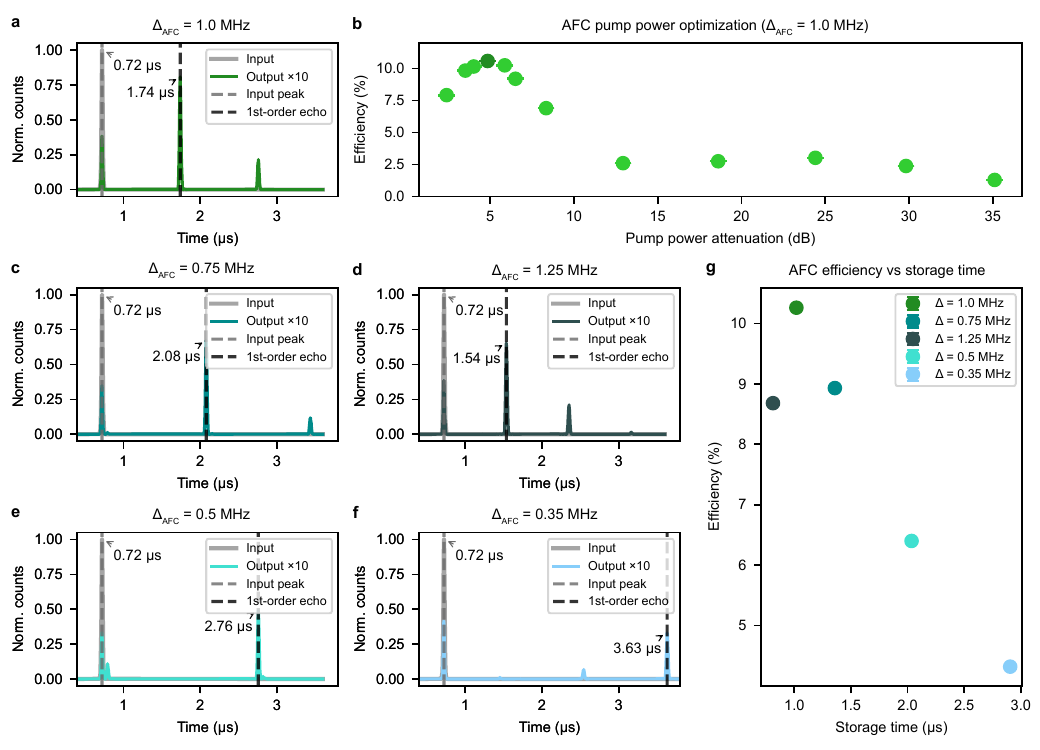}
\caption{\textbf{AFC optimization in $^{166}$Er$^{3+}$:YVO$_{4}$.} \textbf{a}, Storage and retrieval histogram for an optimized AFC with a comb spacing $\Delta_{\mathrm{AFC}}=$~1~MHz. \textbf{b}, Storage efficiency for different AFC preparation (pump) powers. The highest point corresponds to \textbf{a} and is highlighted in dark green. \textbf{c}-\textbf{f}, Storage and retrieval histograms for optimized AFCs with different spacings. Each case is optimized at a different pump power. \textbf{g}, Optimal AFC storage efficiencies for different storage times.}
\label{fig:AFCsummary}
\end{figure*}

For a given AFC teeth spacing, we optimize the AFC storage efficiency ($\eta =$~echo counts/input counts) by varying the pump power, which affects both $d$ and $F$. The input counts are obtained using a pulse far-detuned from the memory resonance. The results depicted in Fig.~\ref{fig:AFCsummary} cover 5 different tooth spacings, 1.25~MHz, 1.0~MHz, 0.75~MHz, 0.5~MHz, and 0.35~MHz. Taking the comb with $\Delta_{\mathrm{AFC}}=$~1.0~MHz spacing as an example, we begin with a low pump power at the order of 10~nW (35~dB attenuation), then gradually increase the power. This ramping not only increases the depth of the teeth (reduces the background level), but also broadens the holes which effectively reduces the linewidth of the teeth and increases the finesse of the comb. When the pump power is too high, however, the broadening of the holes ends up decreasing the contrast and increasing the background level. A full power sweep is shown in Fig.~\ref{fig:AFCsummary}b, where the storage efficiency increases with the pump power until it peaks with an attenuation level of 5~dB, after which the efficiency decreases. Similar power sweeps are repeated for the other 4 combs spacings. In Fig.~\ref{fig:AFCsummary}a, c-f we present the storage histograms (overlaid on top of the off-resonant input pulse) with the optimized pump power in each case. We note that these optimal results are only obtained when we start from a low power level, then gradually ramp up to the optimal power through at least three discrete steps. At each step,  we let the spectral hole to stabilize for a few minutes. If we start directly from a high pump power, the hole will bleach into other neighboring transitions around each pump frequency and prevent us from getting a clean comb spectrum. This is likely due to the specific selection rules between the super-hyperfine levels. The measured AFC storage efficiencies for the five comb spacings are plotted in Fig.~\ref{fig:AFCsummary}g. Based on this, we choose to operate with $\Delta_{\mathrm{AFC}}=$~1.0~MHz for our hybrid networking experiments.

\begin{figure*}[h!]
\centering
\includegraphics{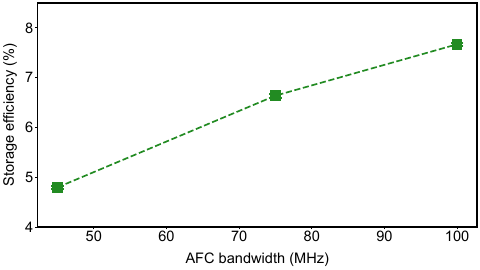}
\caption{\textbf{Dependence of AFC efficiency on bandwidth.} AFC storage efficiencies with different comb bandwidths prepared for an input with a fixed bandwidth of 43~MHz.}
\label{fig:AFCbandwidth}
\end{figure*}

Besides optimizing the optical depth and finesse, we note that Eq.~\ref{eq:AFCefficiency} assumes the input fields are fully absorbed by the comb. This requires the comb bandwidth matched to the bandwidth of the input pulses. In general, the input bandwidth should not exceed the total comb bandwidth. For the measurements in Fig.~\ref{fig:AFCsummary}, the input spectral FWHM is fixed at 24~MHz and the comb bandwidth at 45~MHz. To better approximate the spectrum of the Rb source photon, we increase the weak-pulse FWHM to 43~MHz, which is limited by the AOM rise time. We then characterize the storage using three AFC bandwidths: 45~MHz, 75~MHz, and 100~MHz. As shown in Fig.~\ref{fig:AFCbandwidth}, the significant increase in storage efficiency indicates that, with a broader comb bandwidth, we can approach the optimal efficiency of 10.6(1)$\%$ as shown in Fig.~\ref{fig:AFCsummary}. At present, our maximum bandwidth is constrained by AOMs and by the inhomogeneous broadening (FWHM $\Gamma=131(1)$~MHz) of the crystal. The former limitation could be alleviated using a broadband EOM, whereas the latter can be addressed by co-doping with other rare-earth ions to increase the inhomogeneous linewidth without degrading coherence.

\subsection{Source--Memory coincidence background and temporal gating} \label{SM_gating}

\subsubsection{Modeling gated coincidence background}

From Fig.~\ref{fig:network}b in the main text, we observe a piecewise accidental coincidence background. It consists of four linear functions with respect to the relative delay between the 1530-nm channel and the 780-nm channel. The coincidence between the heralding photon and the non-stored (i.e. directly transmitted) signal photon has a higher background whereas that between the heralding photon and the echo photon has a lower level background. Here we develop a model based on our gating scheme to understand this behavior.

The AOM for optically gating the 1530-nm channel is located in node B (Fig.~\ref{fig:setup}). The arbitrary waveform generator turns it on at $t=0$ for a duration $T_{on}$ ($0 \leq t \leq T_{on}$) and sends a trigger signal at $t=0$ from node B to the Time Tagger in node A for electronically gating the 780-nm channel (Fig.~\ref{fig:control}). The counting channel is activated during $t_{d} \leq t \leq T_{on}+t_{d}$ where $t_{d}$ accounts for the overall delay between the two nodes. Afterwards, both the gating AOM and the 780-nm counting channel are turned off for a duration of $T_{off}$, waiting for the next trigger.

Consider one gating cycle from $t=0$ to $t=T_{on}+T_{off}$ where a photon pair is first generated in node A and the 1530-nm signal photon in the pair reaches the gating AOM (in node B) at $t$. With a different overall delay time of $t_{d}'$, the detection of the un-stored photon at the SNSPD reaches the Time Tagger at $t_{1530}=t+t_{d}'$ and that of the echo reaches there at $t_{1530}=t+t_{d}'+\tau_{\mathrm{AFC}}$.

The detection of all other uncorrelated 780-nm heralding photons is uniformly distributed throughout $t_{d} \leq t_{780} \leq T_{on}+t_{d}$, thus the accidental coincidence at delay $\tau = t_{1530}-t_{780}$ is proportional to the rate of the 1530-nm channel. Now we only consider that from the un-stored 1530-nm photon ($t_{1530}=t+t_{d}'$) which should be uniformly distributed throughout $t+t_{d}'-T_{on}-t_{d}\leq \tau\leq t+t_{d}'-t_{d}$. And we denote the background rate from the source as $b_{h,s}$. We can write the detected accidental coincidence background rate, $R$, as a function of the relative delay $\tau$ between the 780-nm and the 1530-nm channels and the absolute time $t$ when the 1530-nm photon reaches the gating AOM:

\begin{equation}
R(\tau;t) = \begin{cases} 
      b_{h,s} & t+t_{d}'-T_{on}-t_{d}\leq \tau\leq t+t_{d}'-t_{d} \\
     0 & \mathrm{otherwise}
   \end{cases}
\end{equation}
which can be rewritten as:

\begin{equation}
R(\tau;t) = \begin{cases} 
      b_{h,s} & \tau-t_{d}'+t_{d} \leq t\leq \tau-t_{d}'+T_{on}+t_{d} \\
      0 & \mathrm{otherwise}
   \end{cases}
\end{equation}

Then, we can integrate over $0\leq t \leq T_{on}$, for all 1530-nm signal photons that pass through the AOM in this gating cycle, and get the detected background counts, $B_{h,s}$, as a function of the relative delay $\tau$:




\begin{equation}
B_{h,s}(\tau) = b_{h,s}\times\begin{cases} 
      \tau - \tau_{d}+T_{on}& \tau\leq \tau_{d}\\
      -\tau +\tau_{d}+T_{on} & \tau \geq \tau_{d}
   \end{cases}
\end{equation}
where $\tau_{d}=t_{d}'-t_{d}$. Extending to the entire cycle we have:

\begin{equation}
\begin{aligned}
    &B_{h,s}(\tau) =\\
    &b_{h,s}\times
 \begin{cases} 
      \tau - \tau_{d}+T_{on}& 0\leq\tau\leq \tau_{d}\\
      -\tau +\tau_{d}+T_{on} & \tau_{d}\leq \tau \leq \tau_{d}+T_{on}\\
      0& \tau_{d}+T_{on}\leq\tau\leq \tau_{d}+T_{off}\\
      \tau - \tau_{d}-T_{off}& \tau_{d}+T_{off}\leq\tau\leq T_{on}+T_{off}\\
 \end{cases}
\end{aligned}
\end{equation}
The resulting $B_{h,s}(\tau)$ is equivalent to the convolution between two periodic step functions with the same periodicity but a relative shift.

The zero background exists only when $T_{off}>T_{on}$. Therefore, the gating duty cycle has to be less than $50\%$ so that the echo coincidence occurs within the window with suppressed background. Additionally, we need $T_{off}>\tau_{\mathrm{AFC}}$ to optimize the detection for correlated photon pairs. Therefore, the requirement on the gating parameters is $T_{on}<\tau_{\mathrm{AFC}}<T_{off}$. We choose $T_{on}=0.8~\mu$s and $T_{off}=1.2~\mu$s to satisfy this requirement.

\begin{figure*}[h!]
\centering
\includegraphics{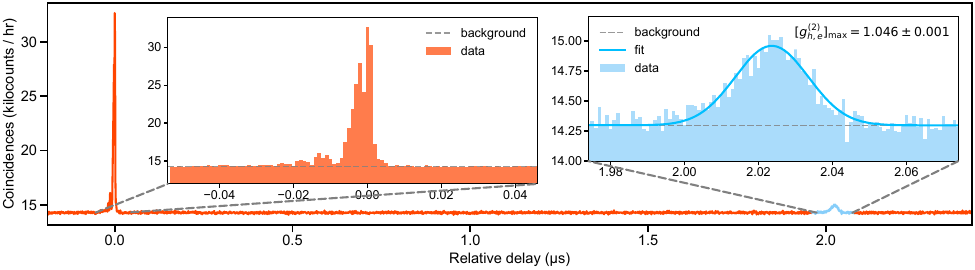}
\caption{\textbf{Source--Memory networking without gating}. Coincidence histogram (bin size: 1~ns) between the 780-nm and the 1530-nm photons after stored and retrieved from the
memory, without implementing the temporal gating scheme. The direct transmission peak (orange) at 0~$\mu$s and the echo (blue) at 2.02~$\mu$s share the same accidental coincidence background level. Note that this dataset is taken with a different source--memory regime from that used in Fig.~\ref{fig:network}.}
\label{fig:gatingcompare_without}
\end{figure*}

\begin{figure*}[h!]
\centering
\includegraphics{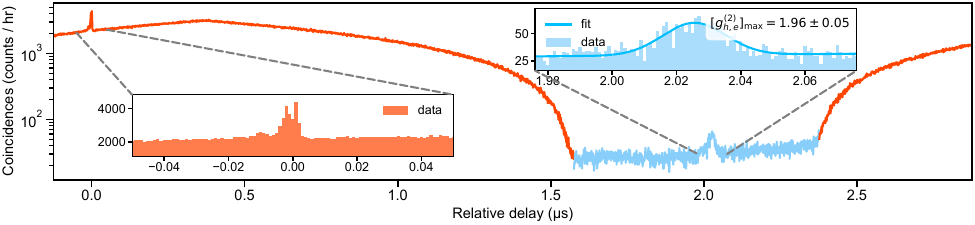}
\caption{\textbf{Source--Memory networking with gating}. Coincidence histogram (bin size: 1~ns) between the 780-nm and the 1530-nm photons after stored and retrieved from the
memory, with the temporal gating scheme implemented. The echo (blue) at 2.02~$\mu$s has a significantly reduced accidental coincidence background level than that of the direct transmission peak (orange) at 0~$\mu$s. This dataset is taken with the same source--memory regime as that used in Fig.~\ref{fig:gatingcompare_without}.}
\label{fig:gatingcompare_with}
\end{figure*}

A comparison between the coincidence histogram without and with gating is provided in Fig.~\ref{fig:gatingcompare_without} and Fig.~\ref{fig:gatingcompare_with}. These two histograms are taken with a source pump regime and an AFC memory configuration marginally different from that used in the main text. Nonetheless, it is clear that the gating significantly improves the cross-correlation $[g^{(2)}_{h,e}]_{\mathrm{max}}$ for the retrieved echo by suppressing the unwanted coincidence background.

\vspace{+20pt}

\subsubsection{Background level characterization}

\begin{figure*}[h!]
\centering
\includegraphics{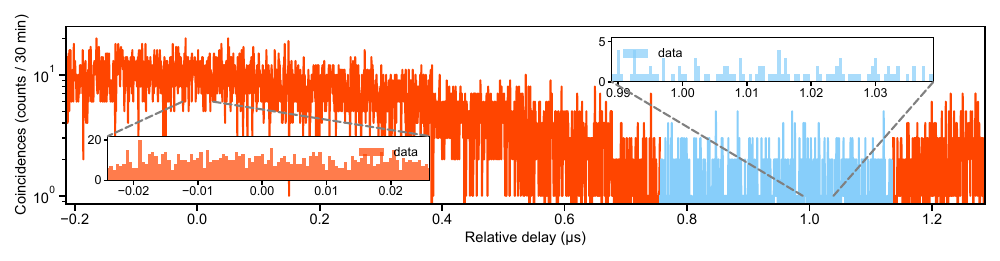}
\caption{\textbf{Accidental coincidences from SNSPD dark counts}. Coincidence histogram (bin size: 0.5~ns) between the 780-nm channel and the 1530-nm channel with the temporal gating scheme implemented, while the collection of 1530-nm photons from the source at node A is blocked, such that no 1530-nm photons are sent to node B. This dataset is taken in the same source–memory regime as that used in Fig.~\ref{fig:network}.}
\label{fig:gating_darkcount}
\end{figure*}

The remaining accidental coincidence background at the retrieved echo consists of three components:
\begin{enumerate}[label=(\roman*)]
\item Coincidences between the uncorrelated 780-nm heralding photons and the retrieved 1530-nm echo photons. This cannot be separated in time from the real coincidence counts between the correlated heralding and echo photons, and thus cannot be gated. However, this contribution can be estimated by multiplying the accidental coincidence background at the source node with the overall storage efficiency.

\item Coincidences between the uncorrelated 780-nm heralding photons and 1530-nm noise photons including detector dark counts. We estimate that this contribution is negligible based on a model described below.

\item Coincidences from SNSPD dark counts. This contribution can be directly measured by blocking all 1530-nm photons from the source node. The result is shown in Fig.~\ref{fig:gating_darkcount}. With the same experiment condition of Fig.~\ref{fig:network}b, we get a coincidence background rate of $D_{\mathrm{SNSPD}}=$~0.52(3)$\times$10$^{-3}$~cps with a 0.5-ns bin.
\end{enumerate}

Here, we derive a simplified model to estimate the noise introduced by the memory node. The model considers: (a) the overall efficiency $\eta$ of the memory; (b) the temporal broadening of the retrieved photons due to internal spectral selection/filtering of the 1530-nm photons from the source; (c) noise added by the AFC memory.

Based on the coincidence histogram between the 780-nm heralding and 1530-nm signal photons from the source, we can express the total coincidence count over all time bins as follows:

\begin{equation}
\label{eq:noise_derive_1}
C_{h,s} =[C_{h,s}]_{\mathrm{max}} \int f_{h,s}(t)\mathrm{d}t
\end{equation}
where $[C_{h,s}]_{\mathrm{max}}$ is the peak coincidence count in one time bin (i.e. the bin that contains the highest count) and $f_{h,s}(t)$ is the normalized temporal envelope of the measured coincidence counts for the source. Likewise, $f_{h,e}(t)$ is that of the echo coincidence. The result of the effect (b) is that the coincidence counts are re-distributed over a larger temporal width, i.e. $f_{h,e}(t)$ has a larger width than $f_{h,s}(t)$.

We then estimate the accidental coincidence count in the 0.5-ns bin after storage and retrieval. Since such background is uniformly distributed over all time bins and not affected by the internal spectral filtering, we have the background for any bin in the echo coincidence histogram as:

\begin{equation}
\label{eq:noise_derive_2}
D_{tot} = \eta\left(\frac{[C_{h,s}]_{\mathrm{max}}}{[g^{(2)}_{h,s}]_{\mathrm{max}}-1}-D_{\mathrm{SNSPD}}\right)+D_{\mathrm{AFC}}+D_{\mathrm{SNSPD}} = \eta\left(\frac{[C_{h,s}]_{\mathrm{max}}}{[g^{(2)}_{h,s}]_{\mathrm{max}}-1}\right)+D_{\mathrm{AFC}}+D_{\mathrm{SNSPD}} ~\mathrm{for} ~\eta \ll 1
\end{equation}
where $[g^{(2)}_{h,s}]_{\mathrm{max}}$ is the maximum cross-correlation that corresponds to the peak coincidence bin. The three terms on the right hand side correspond to the three components of the model described above.

The total echo coincidence count rate over all detection bins can be expressed as:

\begin{equation}
\label{eq:noise_derive_3}
C_{h,e} = [C_{h,e}]_{\mathrm{max}}\int f_{h,e}(t)\mathrm{d}t = ([g^{(2)}_{h,e}]_{\mathrm{max}}-1)D_{tot}\int f_{h,e}(t)\mathrm{d}t
\end{equation}

From our simplified model, we relate the source coincidence count with the echo coincidence count by the overall efficiency:

\begin{equation}
\label{eq:noise_derive}
C_{h,e} = \eta \times C_{h,s}
\end{equation}

Using all raw counts and fitting results from both the source and echo coincidence histograms, we calculate $D_{\mathrm{AFC}}+D_{\mathrm{SNSPD}}=$~0.40(7)$\times$10$^{-3}$~cps. This value is slightly smaller than the independently measured $D_{\mathrm{SNSPD}} =$~0.52(3)$\times$10$^{-3}$~cps, which is likely due to the statistical uncertainty in the SNSPD dark-count coincidence histogram (Fig.~\ref{fig:gating_darkcount}). Nonetheless, the reasonable agreement between the two estimates strongly indicates that the added noise by the AFC memory is negligible.
\vspace{+20pt}

\clearpage

\begin{figure*}[h!]
\centering
\includegraphics{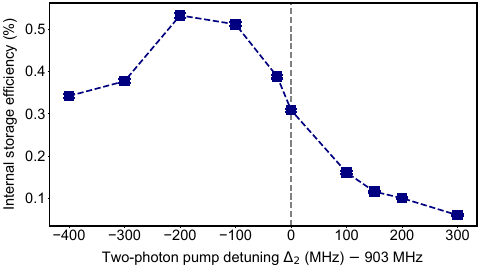}
\caption{\textbf{Internal storage efficiency}. Storage efficiencies for different two-photon pump detunings, calculated by dividing the echo counts shown in the upper-right sub-plot of Fig.~\ref{fig:network}c by the corresponding source count in the upper-left sub-plot, and excluding the measured losses listed in Table~\ref{tab:lossbudget}.}
\label{fig:efficiency}
\end{figure*}

\vspace{+50pt}
\begin{table}[h!]
\centering
\begin{tabular}{lll} \hline
Host (site) & Wavelength (nm) & Frequency (GHz) \\ \hline
Er:Y$_{2}$O$_{3}$ ($C\mathrm{_{3i}}$ site)   & 1545.55 & 193971 \\
Er:MgO & 1540.48 & 194610 \\
Er:Y$_{2}$SiO$_{5}$ (site 2)   & 1538.85 & 194816 \\
Er:Y$_{2}$SiO$_{5}$ (site 1)   & 1536.49 & 195115 \\
Er:Y$_{2}$O$_{3}$  ($C\mathrm{_{2}}$ site)   & 1535.49 & 195242 \\
Er:CaWO$_{4}$ & 1532.63 & 195606 \\
Er:LiNbO$_{4}$ & 1532.00 & 195687 \\
Er:LiYF$_{4}$   & 1530.37 & 195895 \\
Er:GdVO$_{4}$   & 1529.48 & 196010 \\ \hline
Er:YVO$_{4}$   & 1529.21 & 196044 \\
\hline
\end{tabular}
\caption{\textbf{Telecom transitions in Er-doped crystals.} Besides the Er-doped crystals shown above, the transition wavelength in Er-doped fibers is 1532~nm. Rb telecom atomic resonance is around 196038~GHz, making Er:YVO$_{4}$ the closest match.}
\label{tab:hostcrystalsummary}
\end{table}

\vspace{+40pt}
\begin{table}[h!]
\centering
\begin{tabular}{lll} \hline
Component & Source of loss  & $T, \eta (\%)$  \\ \hline
Photon gating & duty cycle & 19 \\
Photon gating & AOM insertion loss & 70 \\
Interconnect & fiber loss & 45 \\
Interconnect & photon collection & 18 \\
Source--Memory & AFC storage & 5\\
Source--Memory  & polarization selection & 50 \\
Source--Memory  & spectral selection & 20 \\
Total & & 0.005\\
\hline
\end{tabular}
\caption{\textbf{System loss budget.} Losses from the photon gating and the interconnect are experimentally measured and constitute external systematic losses; Source–Memory efficiencies constitute the internal storage efficiency.}
\label{tab:lossbudget}
\end{table}

\end{appendices}

\clearpage

\end{document}